\documentstyle[12pt]{article}
\renewcommand{\theequation}{\arabic{section}.\arabic{equation}}
\newcommand{\Or}{{\cal O}}
\newcommand{\C}{{\cal C}}
\newcommand{\CH}{{\cal H}}

\newcommand{\scr}[1]{\mbox{\scriptsize #1}}
\newcommand{\T}{_{\perp}}
\newcommand{\Tt}[1]{_{\perp #1}}
\newcommand{\dd}[2]{\frac{\displaystyle \partial #1}
                         {\displaystyle \partial #2}}
\newcommand{\lsim}{{_{\textstyle <} \atop ^{\textstyle \sim}}}
\newcommand{\beq}{\begin{equation}}
\newcommand{\eeq}{\end{equation}}
\newcommand{\beqa}{\begin{eqnarray}}
\newcommand{\eeqa}{\end{eqnarray}}

\newcommand{\bma}{\left( \begin {array}}
\newcommand{\ema}{\end {array} \right)}
\newcommand{\bfig}{\begin{figure}}
\newcommand{\efig}{\end{figure}}
\newcommand{\bc}{\begin{center}}
\newcommand{\ec}{\end{center}}
\newcommand{\Pslash}{\kern 0.2 em P\kern -0.6em /}
\newcommand{\Phslash}{\kern 0.2 em \hat{P}\kern -0.6em /}
\newcommand{\pslash}{\kern 0.2 em p\kern -0.5em /}
\newcommand{\Khslash}{\kern 0.2 em \hat{K}\kern -0.75em /}
\newcommand{\vslash}{\kern 0.2 em v\kern -0.45em /}
\newcommand{\sla}[1]{\kern 0.2 em #1\kern -0.45em /}
\newcommand{\fra}[2]{\frac{\displaystyle #1}{\displaystyle #2}}
\newcommand{\nummer}[1]{\begin{flushright} #1 \end{flushright}
                        \vspace{-0.2cm}}
\newcommand{\monat}[1]{\vspace{-14pt}\hfill #1
                       \par \vspace*{1 cm}}
\newcommand{\titel}[1]{{\renewcommand{\thefootnote}{\fnsymbol{footnote}}
                       \Large\bf\vskip 0 true cm
                       \begin{center}#1\end{center}
                       \setcounter{footnote}{0}}
                       \normalsize\vskip 1.2 true cm}
\newcommand{\autor}[1]{{
                       \begin{center} {\large #1 }\end{center}}
                       \setcounter{footnote}{0}}
\newcommand{\adresse}[1]{\vspace*{-1.1 true cm}\begin{center} {\it #1 }
                         \end{center}
                         \vskip 0.5cm}

\begin{document}

\thispagestyle{empty}
\nummer{WU-B 93-29 \\ MZ-TH/93-24}
\monat{December 1993}

\titel{ $\ell\!=\!0$ TO $\ell\!=\!1$ TRANSITION FORM FACTORS }
\autor{ J. Bolz
  \footnote{Supported by the Deutsche Forschungsgemeinschaft}
  \footnote{E-mail: bolz@wpts0.physik.uni-wuppertal.de} and
        P. Kroll
  \footnote{Supported in part by the BMFT, FRG under contract 06WU737} }
\adresse{Fachbereich Physik, Universit\"at Wuppertal,\\  D-42097 Wuppertal,
Germany}
\setcounter{footnote}{3}
\autor{J. G. K\"orner~\footnote{Supported in part by the BMFT, FRG
  under contract 06MZ730}}
\adresse{Institut f\"ur Physik, Johannes-Gutenberg-Universit\"at,
  Staudinger Weg 7,\\
  D-55099 Mainz, Germany}

\bc
  {\bf Abstract}
\ec
{\it A method is proposed to extend the hard scattering picture of
Brodsky and Lepage to transitions between hadrons with orbital angular
momentum $\ell\!=\!0$ and $\ell\!=\!1$. The use of covariant spin wave
functions turns out to be very helpful in formulating that method. As
a first application we construct a light-cone wave function of the
nucleon resonance $N^*(1535)$ in the quark-diquark picture. Using this wave
function and the extended hard scattering picture, the $N$--$N^*$
transition form factors are calculated at large momentum transfer and
the results compared to experimental data. As a further application of
our method  we briefly discuss the $\pi$--$\,a_1$ form factors in an
appendix. }

\newpage
\renewcommand{\thefootnote}{\arabic{footnote}}
\setcounter{footnote}{0}

\section{Introduction}

Exclusive processes at large momentum transfer are expected to belong
to the few subjects of hadronic physics amenable to perturbative QCD.
The ``hard scattering picture'' (HSP) \cite{bro80} which was first
formulated by Brodsky and Lepage offers a systematic framework for the
calculation of hadronic quantities like form factors. The basic
assumption of that model is that the physics at small momentum scales
decouples from the hard processes and can be factorized into process
independent distribution amplitudes which are specific to the hadrons
involved. In order to accomplish this factorization the assumption of
collinear constituents has been essential. That assumption is,
however, only reasonable for ground state hadrons. For other hadrons,
having orbital angular momentum $\ell \not= 0$, non-vanishing
transverse momenta ($k\T$) are mandatory. In this paper we attempt a
generalization of the HSP such that also transitions between ground
state hadrons and excited states can be treated. Therefore, the
inclusion of transverse momenta is required which, besides allowing
the investigation of $\ell\!=\!0 \rightarrow \ell\!\not=\!0$ has other
consequences and applications. As was shown by Li and Sterman recently
\cite{lis92} the inclusion of intrinsic transverse momentum helps to
formulate the HSP in a more self-consistent way, in particular it
serves to overcome difficulties connected with the running of the
strong coupling constant $\alpha_S(Q^2)$. It may also allow to study
$\ell\!\not=\!0$ admixtures to ground state hadrons as for instance
the proton. Such admixtures generate hadronic helicity flip
contributions and may therefore explain the experimentally observed
violation of the helicity sum rule of the standard HSP
\cite{bro80}. Thus, one can expect that a consistent treatment of
$k\T$-dependent contributions substantially enriches the HSP.\\
As we are going to demonstrate below the use of covariant spin wave
functions is extremely helpful in formulating the generalization of
the HSP. Covariant spin wave functions for hadrons with internal
orbital angular momentum have first been used by K\"uhn et al.
\cite{kue79,gub80}. These authors constructed spin wave functions via
the $ls$ coupling scheme and expanded both the wave function and the
transition amplitude of a given process in terms of the relative
momentum. These features are also present in the covariant formalism
that we are going to introduce in Sect.~2. \\
In this paper the main interest is focussed on the calculation of the
$N\!\rightarrow N^*(1535)$ transition form factors. In the literature
there are several publications on this issue of which we like to
discuss only a few. First of all Carlson and Poor \cite{car88}
calculated the $N\!\rightarrow N^*(1535)$ transition form factors
within the HSP. They disregarded the transverse momenta of the quarks
and constructed a symmetric distribution amplitude for the $N^*$ the
lowest moments of which were constrained by QCD sum rules. In so far
Carlson and Poor's distribution amplitude is similar to that of the
nucleon, albeit with the opposite parity obtained by modifying the nucleon
covariants by insertion of $\gamma_5$. In a similar manner the same
process has been calculated in the quark-diquark picture \cite{kss92}.
Proceeding in such a manner one obtains the same expressions for the
transition form factors as for the elastic nucleon form factors, up to
kinematical factors. In our opinion such a treatment seems
questionable because according to the $SU(6)$ quark model the $N^*$
spin-flavour function should be of mixed symmetry and thus should differ
markedly from that of the nucleon. Konen and Weber studied the
$N\!\rightarrow N^*$ transition in a relativistic constituent quark
model in the low $Q^2$ regime \cite{kon90}. Although their model is of
limited applicability at large $Q^2$ it contains a correct treatment
of the $N^*$ wave function in the 3-quark picture. For the
nucleon wave function they used the light-cone model of
Dziembowski and Weber \cite{dzw87} and for the $N^*$ they constructed
a proper $SU(6)$ wave function from a totally symmetric Gaussian
momentum distribution function and negative-parity invariants with
an explicit relative momentum dependence.  \\
The goal of this paper is to reanalyze the $N\!\rightarrow N^*$
transition form factors within the HSP but using a proper $N^*$ wave
function now. We will assume the baryons, $N$ and $N^*$, to consist of
a quark and a diquark as in Ref.~\cite{kss92}. The diquark, being a
cluster of two valence quarks and a certain amount of glue and sea
quark pairs, is regarded as a quasi-elementary constituent, which
partly survives medium hard collisions. The composite nature of the
diquark is taken into account by diquark form factors which are
parameterized in such a way that the pure quark picture of Brodsky and
Lepage emerges asymptotically. \\
The diquark picture models non-perturbative effects, in fact
correlations in the baryon wave function, which are known to play an
important r\^ole at moderately large momentum transfer. The
quark-diquark model of baryons has turned out to work rather well for
exclusive reactions. With a common set of parameters specifying the
diquarks and process independent wave functions for the involved
hadrons a good description of a large number of exclusive reactions
has been accomplished by now \cite{kps93,ksP92,kss91}. \\
The paper is organized as follows: In Sect.~2 we lay down the
foundation of how to treat transitions between $\ell\!=\!0$ and
$\ell\!=\!1$ states in the hard scattering picture including a brief review
of the $0\!\rightarrow\!0$ processes considered to date. Sect.~3
is devoted to the formalism of covariant spin wave functions which we will
use for the description of quark-diquark bound states. The calculation
of these form factors is described in Sect.~4 and the results are
presented in Sect.~5. Finally, Sect.~6 is devoted to our summary. The
appendices contain a description of how to construct covariant spin
wave functions using the $ls$ coupling scheme and an examplary
calculation of the $\ell\!=\!0$ to $\ell\!=\!1$ meson transition $\pi
\rightarrow a_1$.

\section{The idea}

We are going to calculate the
$N(\frac{1}{2}^+)\rightarrow N^*(\frac{1}{2}^-)$ transition form
factors at large momentum transfer. For this calculation we use the
hard scattering picture as developed by Brodsky and Lepage
\cite{bro80}, but generalized in such a way that also hadrons with
non-zero orbital angular momentum between their constituents can be
treated. The main idea of that generalization is presented in this
section; its explicit application to the $N$--$N^*$ transitions will
be discussed in Section 5. The approach we present here may also be
applied to other hadron-hadron transitions with $\ell \not= 0$. \\
As usual we use the light-cone approach, or --- equivalently --- infinite
momentum frame techniques, to formulate the generalized hard
scattering picture. In the light-cone approach, which is a covariant
framework for describing a composite system with a fixed number of
constituents, a hadron is described by a momentum-space Fock basis.
The coefficients of the various states of that basis represent wave
functions defined at equal light-cone time $\tau\,=\,t+z$, rather than
at equal $t$. Since the light-cone approach enables one to completely
separate the kinematical and dynamical features of the Poincar\'e
invariance \cite{dir49,leu78}, there is a factorization of the basis
hadronic states into the kinematical and dynamical parts, i.e. the
overall motion of the hadron is decoupled from the internal motion of
the constituents. \\
Supppose the hadron's momentum is $P = (P^+,P^-,\vec P\T)\:$ where
$P^+=P\,^0 + P\,^3$ and $P^-=(P\,^0 - P\,^3)/2$. To accomplish the
factorization into overall and internal motion, the momentum of the
$j$-th parton belonging to a given Fock state is characterized by the
fraction $x_j=p_j^+/P^+$ and by the transverse momentum $\vec p\Tt{j}
= x_j \vec P\T + \vec k\Tt{j}$. Momentum conservation provides two
constraints on the parton momenta of that Fock state
\beq
   \sum_{j=1}^n x_j \:=\: 1 \:; \qquad \qquad
   \sum_{j=1}^n \vec k\Tt{j} \:=\: 0 \:.
   \label{partcon}
\eeq
It can be shown \cite{bro80,leu78,dzi88} that the variables $x_j$ and
$\vec k\Tt{j}$ are invariant under all kinematical Poincar\'e
transformations, i.e. under boosts along and rotations around the
3-direction as well as under transverse boosts. Moreover --- and this
is the essential point --- the light-cone wave function $\psi \,=\,
\psi(x_j,\vec k\Tt{j})$ for that Fock state is independent of the
hadron's momentum and is invariant under these kinematical
transformations too. Hence, $\psi$ is determined if it is known at
rest.
An ordinary equal-time wave function does not possess these
properties. The wave function $\psi$ may depend on internal quantum
numbers such as helicities, a possibility which we omit for
the present discussion. We emphasize that only the spin operator $J_3$
is purely kinematical, whereas the other two spin operators depend on
the interaction. Consequently, rotational invariance is difficult to
implement in the light-cone formalism.
As long as only hadrons with zero orbital angular momenta between
their constituents are considered, this difficulty disappears
\cite{bro80}. This is also the case for the form factors of
transitions between $\ell\!=\!0$ and $\ell\!=\!1$ hadrons as we
are going to show below. In general, however, this is not true. Thus,
for instance, for exclusive decays of the P-wave charmonia a
consistent, rotational invariant light-cone approach has not yet been
achieved. In order to preserve rotational invariance one usually
describes the charmonium state by an ordinary equal-$t$ wave function,
whereas light-cone wave functions are used for the final state hadron
\cite{che89}. \\
It can be shown \cite{bro80} that the contributions from the valence
Fock state dominate at large momentum transfer, higher Fock state
contributions are suppressed by powers of $\alpha_S/Q^2$. Therefore,
one customarily assumes valence Fock state dominance in applications
of the hard scattering picture, although, at moderately large momentum
transfer --- the region where data is available --- it is by no
means obvious that contributions from higher Fock states are indeed
negligible \cite{rad91,jak93}. The valence Fock state dominance is
particularly questionable for $\ell\!\not=\!0$-hadrons \cite{bod92}.
Nevertheless, we will make use of this assumption. \\
Let us now discuss the main features of the $N\!-\!N^*$ transition
form factor calculation at large or moderately
large momentum transfer. Because of the reasons mentioned in the
introduction we regard a baryon as a bound state of a quark and a
diquark where the latter is treated as a quasi-elementary constituent
(for details, see Sects.~3 and 5). We perform the calculation in a
frame in which both baryons move along the 3-direction. In particular
we use the opposite momentum brick wall frame defined by
$\hat P_i^{\mu}=(E_i-P,\,(E_i+P)/2,\,\vec 0\T)$ and
$\hat P_f^{\mu}=(E_f+P,\,(E_f-P)/2,\,\vec 0\T)$ where
$\hat P_j^{\mu}$ and $\hat P_f^{\mu}$ denote the momenta of the
initial and final state baryon, respectively. Any other frame that is
obtained by a boost along the 3-direction from the brick wall frame
belongs to the set of frames in the 3-direction, hence the infinite
momentum frame too.  \\
In order to bring out the main features for the moment we ignore all
complications in connection with flavour wave functions as well as
the possibility that several different quark-diquark configurations
may contribute to a given baryon. We make the following ansatz for a
spin $\frac{1}{2}$-baryon of type $a$
\footnote{ Note that, for conciseness, we have omitted colour indices
and a factor representing the plane waves. }
\beq
  |\: a;\,\vec P_a, \, \lambda_a \, \rangle \:=\:
      \int\, \frac{dx_1\,d^2k\T}{16 \pi^3} \:
      \psi_a(x_1,\vec k\T)\:\Gamma_a\:u_a(\vec P_a,\lambda_a) \, ,
  \label{barans}
\eeq
where $P_a$ and $\lambda_a$ are the momentum and the helicity of the
baryon; $u_a$ represents its light-cone spinor. It is
normalized as $\bar u u \,=\, 1$. The variables $x_1$ and $\vec k\T$
refer to the quark; the diquark variables are $x_2=1-x_1$ and $-\vec
k\T$. $\Gamma_a$ represents the covariant spin
wave function of the baryon \cite{hus91,bhk93}.
Here, in this section, we only need a few basic features of the spin
wave functions; their explicit forms for the baryons under
consideration will be specified in Section 3.
A spin wave function depends on the hadronic spin and parity as well
as on the quantum numbers of the constituents. But, with the
exception of $k\T$, it does not depend on the momenta or
polarisation vectors (helicities) of the constituents. This fact
entails an enormous technical advantage in the calculation of
matrix elements. One of the basic ingredients of the hard scattering
picture is the so-called collinear approximation which says that all
constituents move along the same direction as their parent hadron up
to a scale of the order of the Fermi momentum $<k\T^2>^{1/2}$ which is
typically a few 100 MeV. This collinear approximation justifies an
expansion of the spin wave function in terms of a power series in
$\vec k\T$ or, in order to retain the covariant formulation, in
$K^{\mu}\,=\,(\,K^+\!=\!0,\,K^-\!=\!0,\,\vec k\T\,)$. Up
to terms linear in $K^{\mu}$ this expansion reads
\beq
  \Gamma_a\,(\ell\!=\!0) \:=\: \Gamma_{a0}(K\!=\!0) \,+\,
        \Delta\Gamma_{a0}^{\alpha}(K\!=\!0) \, K_{\alpha} \,+\, \Or(k\T^2) \, ,
  \label{exp0}
\eeq
for a $\ell\!=\!0$ baryon. The light-cone wave function $\psi_a$ of
the baryon $a$ is a scalar function in this
case; it depends only on $k\T^2$, not on $\vec k\T$. \\
Contrary to the above ansatz, the momentum
$K^{\mu}=(p_1^{\mu}\!-\!p_2^{\mu})/2$  (see Sect.~3) possesses a
non-vanishing $K^- (=(K^0\!-\!K^3)/2)$-component in general. Writing
the constituent momenta in a covariant fashion as ($j=1, 2$)
\beq
  p_j^{\mu} \:=\: x_j P^{\mu} \,+\, k_j^{\mu} \:,
  \label{pj}
\eeq
where, in agreement with the required invariance under the kinematical
Poincar\'e transformations,
\beq
  k_1^{\mu} \:=\: (\,0,\,k_1^-,\,\vec k\T\,)\:, \qquad \qquad
  k_2^{\mu} \:=\: (\,0,\,k_2^-,\,-\vec k\T\,)\:.
  \label{kj}
\eeq
For the minus-component of the relative momentum we have
\beq
  K^- \:=\: \frac{1}{2} (k_1^- - k_2^-) \, .
  \label{kminus}
\eeq
As is customary in the parton model, we neglect the binding energy and
consider the constituents as on-shell particles. That possibly crude
approximation can be achieved by putting the individual
$k_j^-$-components to zero. With this choice also $K^-$ is zero and
hence $P\!\cdot\!K = 0$. As we will see this choice leads to a very
elegant and compact representation of the transition matrix elements.
Under these circumstances the constituent velocities equal that of
their parent hadron up to corrections of order $k\T$. \\
In the case of non-zero orbital angular momentum between the two
constituents of a given baryon a non-zero transverse momentum
is required. Indeed one can easily show \cite{hus91} that in the case
of $\ell=1$ the following expansion holds for the spin wave function
\beq
  \Gamma_a(\ell\!=\!1) \:=\: \Gamma_{a1}^{\alpha}(K\!=\!0) \, K_{\alpha} \,+\,
        \Delta\Gamma_{a1}^{\alpha\beta}(K\!=\!0) \, K_{\alpha} K_{\beta}
     \,+\, \Or(k\T^3) \, .
  \label{exp1}
\eeq
$\psi_a$ in (\ref{barans}) now represents a reduced wave function,
i.e. the full wave function with a factor $K^{\mu}$ removed from it.
In the non-relativistic case (see Appendix A) the factor $K^{\mu}$
arises from the spherical function and $\psi_a$ is related to the
radial function. As in the $\ell=0$ case the reduced wave function
$\psi_a$ depends only on $\vec k\T^2$. \\
In both cases, $\ell\!=\!0$ and $\ell\!=\!1$, we normalize the
spin wave functions in such a way that
\beq
  \bar u_a(P_a, \lambda_a') \, \bar \Gamma_a \, \Gamma_a \,
    u_a(P_a,\lambda_a) \:=\: k\T^{2\ell} \, \delta_{\lambda_a \lambda_a'}
    \:+\: \Or(k\T^{2\ell+2}) \, ,
  \label{wnorm}
\eeq
where, as usual, $\bar \Gamma_a \,=\, \gamma_0 \, \Gamma_a^{\dag} \,
\gamma_0$. Proper state normalization requires the condition
\beq
  \int\, \frac{dx_1\,d^2k\T}{16 \pi^3} \;
      |\,k\T^{\ell}\,\psi_a(x_1,k\T)\,|^2 \:=\:\kappa_a \:\leq\: 1 \, .
  \label{normcond}
\eeq
$\kappa_a$ is to be interpreted as the probability of finding the
given quark-diquark configuration inside the baryon $a$. \\
Within the hard scattering picture the current matrix elements for the
transitions from a baryon $i$ to a baryon $f$ are represented by
convolutions of the corresponding wave functions and hard scattering
amplitudes $T_H$. The latter are to be calculated perturbatively from an
appropriate set of Feynman diagrams:
\beqa
  \lefteqn{
      \langle\,f; \, \vec P_f, \, \lambda_f \:|\:J^{\mu}\:|\: i;\,\vec P_i,\,
      \lambda_i \, \rangle \:=\: \bar u_f(P_f,\lambda_f) } \nonumber \\
  &\times& \int\, \fra{dx_1\,d^2k\T\,dy_1\,d^2k'\T}
      {(16 \pi^3)^2} \: \psi_f^*(y_1,k'\T)\:
      \bar\Gamma_f\:T_H^{\mu}(x_1,y_1,Q,K,K')\:\Gamma_i\:\psi_i(x_1,k\T)\:
      u_i(P_i,\lambda_i) \, . \hspace{1cm}
 \label{curmat}
\eeqa
$Q^2\:(\ge0)$ is the usual invariant momentum transfer from the initial
to the final state baryon. For the final baryon the momentum
fractions, the transverse momentum and the relative momentum are
denoted by $y_j$, $k'\T$ and $K'$ respectively. Comparison of
(\ref{curmat}) with a standard decomposition of the current matrix
elements in terms of covariants and invariant form factors leads to a
determination of the form factors within the hard scattering model. \\
At large $Q^2$ the intrinsic transverse momentum dependence of the hard
scattering amplitude provides only a higher twist correction. In view
of this, we also expand the amplitude in a power series of the
transverse momenta:
\beqa
  \lefteqn{T_H^{\mu}\,(x_1,y_1,Q,K,K') \:=\:} \nonumber \\
 & &  T_H^{\mu}(x_1,y_1,Q,K\!=\!K'\!=\!0)
      \,+\,\dd{T_H^{\mu}}{K^{\alpha}} \bigg|_{K=K'=0} K^{\alpha}
      \,+\,\dd{T_H^{\mu}}{K'^{\alpha}} \bigg|_{K=K'=0} K'^{\alpha}
      \,+\, \Or(k\T^2,k\T^{\prime 2}) \, . \hspace{1cm}
  \label{THexp}
\eeqa
Let us now demonstrate that on keeping only the first term on the
r.h.s.~of (\ref{THexp}) one recovers the usual $i(\ell\!=\!0)
\rightarrow f(\ell'\!=\!0)$ hard scattering formula.
Inserting (\ref{THexp}) and (\ref{exp0}) or (\ref{exp1}) into
(\ref{curmat}) and noting that
\beq
  \int\, \frac{dx_1\,d^2k\T^{(\prime)}}{16 \pi^3} \;
      K^{(\prime)\mu}\,\psi_{i(f)}(x_1,k\T)\:=\:0 \:,
\eeq
we find
\beqa
   \lefteqn{
      \langle f(\ell'\!=\!0);\,\vec P_f,\,\lambda_f \:|\:J^{\mu}\:|
      \:i(\ell\!=\!0);\,\vec P_i, \,\lambda_i \, \rangle \:=\:
      \bar u_f(P_f,\lambda_f)\,
      \int \frac{dx_1 d^2k\T dy_1 d^2k'\T}{(16 \pi^3)^2} \,
      \psi_f^*(y_1,k'\T) } \hspace{2cm} \nonumber \\
  & \times &
      \bar\Gamma_{f0}(K'\!=\!0)\,
      T_H^{\mu}(x_1,y_1,Q,K\!=\!K'\!=\!0) \,
      \Gamma_{i0}(K\!=\!0)\,\psi_i(x_1,k\T)\,
      u_i(P_i,\lambda_i)  \nonumber \\
  & + &  \Or(k\T^2,k\T^{\prime 2}) \, . \hspace{11cm}
  \label{0to0}
\eeqa
The integrations over the transverse momenta apply only to the wave
functions. Carrying out these integrations formally, one arrives at
the so-called distribution amplitudes (DA) ($a=i,f$).
\beq
   f_a\:\phi_a(x_1) \:=\: \int\, \frac{d^2k\T}{16\pi^3}
                                \:\psi_a(x_1,k\T) \, ,
   \label{da0}
\eeq
where, by convention, the distribution amplitudes are defined in such
a way that
\beq
   \int\, dx_1 \: \phi_a(x_1) \:=\: 1 \, .
   \label{normda0}
\eeq
We have factored out the coupling factor $f_a$ which represents the
configuration space wave function at the origin. With these
defintions, Eq.~(\ref{0to0}) simplifies to
\beqa
   \lefteqn{
      \langle\,f(\ell'\!=\!0);\,\vec P_f,\,\lambda_f \:|\:J^{\mu}\:|
      \:i(\ell\!=\!0);\,\vec P_i, \,\lambda_i \, \rangle \:=\: }
      \hspace{2cm} \nonumber \\
 & &  \bar u_f(P_f,\lambda_f) \,\int dx_1 dy_1\,f_f\,\phi_f^*(y_1)\,
      \bar\Gamma_{f0}(K'\!=\!0)\,T_H^{\mu}(x_1,y_1,Q,K\!=\!K'\!=\!0)\,
      \nonumber \\
 & \times &  \Gamma_{i0}(K\!=\!0)\,
      f_i\,\phi_i(x_1)\:u_i(P_i,\lambda_i) \:+\:
      \Or(k\T^2,k\T^{\prime 2}) \, . \hspace{5cm}
  \label{0to0da}
\eeqa
This is the standard hard scattering formula as derived by Brodsky and
Lepage \cite{bro80}. The integrand factorizes into long-distance
physics, contained in the distribution amplitudes, and
short-distance physics, contained in the hard scattering amplitudes.
In principle the distribution amplitudes depend logarithmically on Q
as a consequence of the QCD evolution \cite{bro80} which will,
however, not be taken into account in our calculation. \\
For the case $i(\ell\!=\!0) \rightarrow f(\ell'\!=\!1)$, on the other hand,
we find
\beqa
  \lefteqn{
      \langle\,f(\ell'\!=\!1);\,\vec P_f,\,\lambda_f \:|\:J^{\mu}\:|
      \:i(\ell\!=\!0);\,\vec P_i,\,\lambda_i \, \rangle \:=
      \bar u_f(P_f,\lambda_f) \,
      \int\, \frac{dx_1 d^2k\T dy_1 d^2k'\T}{(16 \pi^3)^2}
      \psi_f^*(y_1,k'\T)\,} \hspace{2cm}  \nonumber \\
  &\times& K'_{\alpha} K'_{\beta}\, \Bigg[\,
      \bar\Gamma_{f1}^{\alpha}(K'\!=\!0)\,\dd{T_H^{\mu}}{K'_{\beta}}
      \bigg|_{K=K'=0} \,+\, \Delta\bar\Gamma_{f1}^{\alpha\beta}
      (K'\!=\!0)\,T_H^{\mu}(K\!=\!K'\!=\!0) \Bigg]\nonumber \\
  &\times& \Gamma_{i0}(K\!=\!0)\,
      \psi_i(x_1,k\T)\,u_i(P_i,\lambda_i)\:+\:\Or(k\T^2 k\T^{\prime 2},
      k\T^{\prime 4}) \, , \hspace{5cm}
  \label{0to1}
\eeqa
to the requisite order of $k\T$. Note that terms linear in
$K_{\alpha}$ or $K'_{\alpha}$ drop out after integration. One notes
that the first order correction term to the $\ell'\!=\!1$-spin wave
function enters the matrix element to the same order as the leading
wave function term. Thus, in contrast to the $\ell\!=\!0$-case, the
correction term to the $\ell'\!=\!1$-spin wave function is required at
leading order.~\footnote{A process for which this term is of
particular relevance is  the two-photon decay of the $\chi_0$. The
term $\sim \Gamma_1$ is zero. Hence, the familiar result for the decay
width \cite{bar76} is solely obtained from the term $\sim
\Delta\Gamma_1$.}
Again, as in the $\ell\!=\!0$-case, the transverse momentum integrations
apply only to the wave functions. With the aid of the covariant
integration formula
\beq
   \int\, \frac{d^2k'\T}{16\pi^3}\:K'_{\alpha} K'_{\beta} \:
	\psi_f(y_1,k'\T) \:=\: f_f \: \phi_f(y_1) \: I_{\alpha\beta}
        / 2  \:,
   \label{intform}
\eeq
the tensor integration in (\ref{intform}) can be done explicitly with
the result
\footnote{The metric tensor reads in light-cone coordinates
\[ g_{\alpha\beta} = \bma{cccc} 0 & 1 &  0 &  0 \\
                                1 & 0 &  0 &  0 \\
                                0 & 0 & -1 &  0 \\
                                0 & 0 &  0 & -1 \ema \, . \]
The scalar product of two 4-vectors is
$a\cdot b = a^+b^- + a^-b^+ -\vec a \cdot \vec b$.
}
\beq
   I_{\alpha\beta} \:=\: -g_{\alpha\beta} \,+\, \frac{1}{M_f^2}
          (P_{f\alpha}P_{f\beta}\,-\,\bar P_{f\alpha}\bar P_{f\beta})
          \, .
   \label{iab}
\eeq
The momentum $\bar P_{f\alpha}$ is given by $(\,P_f^+,\,-P_f^-,\,\vec
0\T\,)$ in the frame we are working. The remaining scalar integral
serves to define a distribution amplitude in analogy to the
$\ell\!=\!0$ case, as mentioned before. One has
\beq
   f_f\:\phi_a(y_1) \:=\: \int\, \frac{d^2k'\T}{16\pi^3}
                           \:k\T^{\prime 2}\:\psi_f(y_1,k'\T) \, ,
   \label{da1}
\eeq
where
\beq
   \int\, dy_1 \: \phi_f(y_1) \:=\: 1 \, .
   \label{normda1}
\eeq
Again we have factored out a coupling factor $f_f$ which is to be
interpreted as the derivative of the configuration space wave function
with respect to $r$ at the origin. \\
Inserting (\ref{intform}-\ref{normda1}) into (\ref{0to1}) we obtain
\beqa
  \lefteqn{
      \langle\,f(\ell'\!=\!1);\,\vec P_f,\,\lambda_f \:|\:J^{\mu}\:|
      \:i(\ell\!=\!0);\,\vec P_i,\,\lambda_i \, \rangle \:=
      \bar u_f(P_f,\lambda_f) \, \int dx_1 dy_1 \,}
      \nonumber \\
  & \times & \frac{1}{2} f_f \,\phi_f^*(y_1)\,I_{\alpha\beta}
      \,\Bigg[ \,\bar\Gamma_{f1}^{\alpha}(K'\!=\!0)\,
      \dd{T_H^{\mu}}{K'_{\beta}}\bigg|_{K=K'=0}
      \,+\,\Delta\bar\Gamma_{f1}^{\alpha\beta}(K'\!=\!0)\,
      T_H^{\mu}(K\!=\!K'\!=\!0) \Bigg] \nonumber \\
  & \times &   \Gamma_{i0}(K\!=\!0)\,
      f_i\,\phi_i(x_1)\,\,u_i(P_i,\lambda_i)\:+\:\Or(k\T^2 k\T^{\prime 2},
      k\T^{\prime 4}) \, . \hspace{5cm}
  \label{0to1da}
\eeqa
This is the generalized hard scattering formula. We will use it to
calculate the $N\!\rightarrow\!N^*$-transition form factors (see
Sect.~4). \\
The following remarks are in order:
\newcounter{li}
\begin{list}{\roman{li})}{\usecounter{li}
  \labelwidth0.5cm \leftmargin0.7cm \labelsep0.2cm \rightmargin0cm
  \topsep-0.2cm \itemsep0cm}
\item Eq.~(\ref{0to1da}) factorizes into long and short distance
physics. As in the standard hard scattering formula (\ref{0to0da}) the
long distance physics is contained in the distribution amplitudes, the
short distance physics in the hard scattering amplitude and its
derivative.
 \item The generalization of (\ref{0to1da}) to
$\ell\!\rightarrow\!\ell'$-transitions for any values of $\ell$ and
$\ell'$ is straightforward.
 \item We have presented the generalized hard scattering formula for
baryonic transitions. The same approach, however, also holds for
mesonic transitions. One only has to replace the baryon spinors in
(\ref{0to1da}) by a trace over Dirac matrices. In Appendix B we
demonstrate this for the $\pi \rightarrow a_1$ transitions.
 \item We have used the quark-diquark model for the baryons because of
our actual phenomenological interest. The approach to
$\ell\!\rightarrow\!\ell'$-transitions that we presented in this
section can easily be adapted to the pure quark picture of baryons.
 \item Recently, Sterman and Li \cite{lis92} have modified the hard
scattering picture (for $\ell\!=\!0 \rightarrow \ell\!=\!0$
tran\-sitions). They kept
the transverse momentum dependence of the hard scattering amplitude
and took into account radiative (Sudakov) corrections. Because of that
the perturbative QCD contribution to form factors becomes
self-consistent even for momentum transfers as low as a few GeV. Here
self-consistency means that most of the result comes from hard regions
where $\alpha_S$, the strong coupling constant, is small. As has been
pointed out in \cite{jak93} the consistency of the entire approach
also requires the inclusion of the intrinsic transverse momentum
dependence of the hadronic wave function. The approach put forward by
Sterman and Li challenges previous objections \cite{rad91,isg89} against
the use of the hard scattering picture.
Our approach to $\ell\!\rightarrow\!\ell'$-transitions can be generalized
such as to include Sudakov corrections as well. However, we dispense
with these corrections in this explorative attempt to calculate
$\ell\!=\!0 \rightarrow \ell\!=\!1$ transitions.
\end{list}


\section{The covariant spin wave functions for the nucleon and the
         $S_{11}$-nucleon resonance}

\setcounter{equation}{0}
We now discuss in some detail the wave functions for the two baryons
of interest. As we have mentioned repeatedly we consider baryons to be
bound states of quarks and diquarks. We emphasize that for each
baryon under consideration the spin-flavour wave functions are
constructed in such a way that after resolving the diquarks into two
quarks the corresponding representation of the static $SU(6)$-model
emerges. This requirement puts constraints on possible quark-diquark
configurations inside a baryon as well as on relative factors between
the different terms in the quark-diquark wave function. \\
Covariant baryon spin wave functions have been derived in a very
general way in Ref.~\cite{hus91}. They are presented in that paper in
a way such that the spin wave functions can easily be interpreted as
originating from a quark-diquark picture although in Ref.~\cite{hus91}
this picture itself has not been used. \\
The introduction of covariant spin wave functions allows one to
covariantly calculate the probability amplitude that a given spin
configuration of the quark and diquark constituents is contained in a
given baryon of any spin and parity. In this sense the covariant spin
wave functions represent the Clebsch-Gordan coefficients of a
particular rest frame spin coupling scheme boosted to the moving
frame and can be derived from the standard non-relativistic
$ls$-coupling scheme (see Appendix A). \\
In the rest frame of the baryon, where the velocity 4-vector
takes the form $v_a^{\mu}=P_a^{\mu}/M_a=(1,\vec 0)$, the covariant spin
coupling reduces to couplings of spin objects transforming under the
three dimensional rotation group $O(3)$ or $SU(2)$. In order to
achieve this, one has to assure that the time or spin zero components
of Lorentz tensors that are used in the construction of the covariant
spin wave functions decouple in the rest frame. One therefore has to
use 4-transverse angular momentum vectors and tensors in the
construction of the covariant spin wave functions. For example
when $K^{\mu}$ is the relative momentum
\beq
   K^{\mu} \:=\: (p_1^{\mu} - p_2^{\mu}) / 2 \, ,
   \label{krel}
\eeq
each unit of orbital momentum will be represented by
\beq
   K\Tt{a}^{\mu} \:=\: K^{\mu} - v_a\!\cdot\!K\,v_a^{\mu} \, .
   \label{ktr}
\eeq
In the rest frame clearly $K\T^{\mu} \rightarrow (0,\vec k)$ and one has
the appropriate object transforming as a 3-vector under $O(3)$.
Considering the fact that the spin wave functions act on baryon spinors
one can easily convince oneself that the relative momentum is only
determined up to a multiple of the hadron momentum. Combining this
freedom with the zero binding energy approximation, one can define $K$
as we did in Sect.~2, namely in such a way that it has only the
transverse components $\vec k\T$. This choice, although not forced, is
very convenient. \\
Similarly, for a baryon of type $a$, one has to work with the
4-transverse Dirac matrix
\beq
  \gamma\Tt{a}^{\mu} \:=\: \gamma^{\mu} \,-\, v_a^{\mu} \vslash_a
                 \:=\: \left(\,g^{\mu\nu}\,-\,v_a^{\mu}v_a^{\nu}
                       \right)\,\gamma_{\nu}\, ,
  \label{gamtr}
\eeq
which clearly reduces to $(0,\vec \gamma)$ in the rest frame of the
baryon representing a spin operator with the correct transformation
property under $O(3)$. We mention that only one of the off-diagonal
$2\times 2$ matrices in the rest frame form of $\gamma\Tt{a}^{\mu}$ is
of relevance since we are working with fermions (and not
anti-fermions). \\
In writing down the baryon states we have to consider the possibility
that a baryon may be composed of several quark-diquark configurations.
The ground state nucleon is made up of two configurations when
admixtures from configurations with $\ell\not=0$ are neglected. In terms
of the non-relativistic quark-model this means that we assume the
nucleon to be a pure state of the $SU(6)$ $\{56\}$ multiplet. The two
configurations of the nucleon consist of a quark and either a spin 0,
isospin 0 diquark ($S$) or a spin 1, isospin 1 diquark ($V$).
According to Sect.~2, in particular Eqs.
(\ref{barans}, \ref{exp0}, \ref{da0}), we therefore write a nucleon
state as
\beq
   |\:N;\,\vec P_i,\,\lambda_i\,\rangle \:=\: \int dx_1\: \left[ \, f_{NS} \,
       \phi_{NS}(x_1)\,\chi_{NS}\,\Gamma_{NS} \:+\: f_{NV} \,
       \phi_{NV}(x_1)\,\chi_{NV}\,\Gamma_{NV}^{\mu} \, \right] \:
       u_N(P_i,\lambda_i) \, ,
   \label{nucstate}
\eeq
where the covariant spin wave functions are given by
\beq
   \Gamma_{NS}\:=\:1 \, ; \qquad\qquad
   \Gamma_{NV}^{\mu} \:=\:  \frac{1}{\sqrt{3}}\,\gamma\Tt{i}^{\mu}\,
         \gamma_5    \:=\:  \frac{1}{\sqrt{3}}\,(\gamma^{\mu}+v_i^{\mu})\,
         \gamma_5 \, .
   \label{gamnuc}
\eeq
In (\ref{gamnuc}) we have made use of the fact that the spin wave
function acts on the baryon spinor, i.e. we have replaced $\vslash_i$
by 1.The spin wave functions are normalized according to (\ref{wnorm}).
We do not need $K$-dependent correction terms to these spin wave
functions to the order we are working (see (\ref{0to0})). The $\chi$'s
in Eq.~(\ref{nucstate}) represent flavour functions, which, for the
proton and neutron, take the form
\beq
  \begin{array}{l}
    \chi_{pS} \:=\: u\,D^0_{[u,d]} \, , \\
    \chi_{nS} \:=\: d\,D^0_{[u,d]} \, ,
  \end{array}  \qquad\qquad
  \begin{array}{l}
    \chi_{pV} \:=\: [u\,D^1_{\{u,d\}} - \sqrt{2}\,d\,D^1_{\{u,u\}} ]
                    / \sqrt{3} \: , \\
    \chi_{nV} \:=\: -[d\,D^1_{\{u,d\}} - \sqrt{2}\,u\,D^1_{\{d,d\}}]
                    / \sqrt{3} \: ,
  \end{array}
  \label{chinuc}
\eeq
where $D^0_{[u,d]}(D^1_{\{u,d\}})$ stands for an isospin 0 (1) diquark
made of a $u$ and a $d$ quark. \\
The form (\ref{nucstate}) of the nucleon state has been used in many
applications of the hard scattering picture. Thus, for instance, the
electromagnetic form factors of the nucleon have been investigated in
both the space-like and the time-like regions
\cite{kps93}. Also Compton-scattering, two photon annihilation into
proton-antiproton, photoproduction of mesons as well as some exclusive
decay processes have been studied. For a recent review of applications
of the quark-diquark picture, see \cite{ksP92}. All of these studies have
been carried out with the following distribution amplitudes
\beqa
  \label{phiNS}
  \phi_{NS}(x_1) &\:=\:& \C_{NS} x_1 x_2^3 \exp \left[ -\beta_N^2 (m_q^2/x_1 +
                        m_S^2/x_2)\right] \, ,\\
  \phi_{NV}(x_1) &\:=\:& \C_{NV} x_1 x_2^3 (1 + 5.8 x_1 - 12.5 x_1^2)
                \exp \left[ -\beta_N^2 (m_q^2/x_1 + m_V^2/x_2)\right] \, .
  \label{phiNV}
\eeqa
We reiterate that $x_1$ and $x_2 = 1-x_1$ refer to the longitudinal
fractions of the quark and the diquark, respectively. The DAs
represent a kind of harmonic oscillator wave function transformed to
the light-cone. The masses in the exponentials are, therefore,
constituent masses since they enter through a rest frame wave
function. For the quark mass we take 330 MeV, whereas for the diquark
masses a value of 580 MeV is used. The oscillator parameter $\beta_N$
is taken to be 0.5 GeV$^{-1}$. This value is obtained by the
requirement that the root mean square (rms) transverse momentum
$<k\T^2>^{1/2}$ is 600 MeV (as found e.g. by the EMC \cite{EMC80} in a
study of transverse momentum distributions in semi-inclusive deep
inelastic $\mu$--$p$ scattering), assuming a Gaussian $k\T$ dependence
for the full wave function
\beq
   \sim\:\exp\left[\,-\beta_N^2\,\frac{k\T^2}{x_1 x_2}\,\right] \, .
   \label{ktdep}
\eeq
The exact values of the masses and that of the oscillator parameter
are not very important since the exponentials in (\ref{phiNS},
\ref{phiNV}) are only of importance in the end-point regions. The
constants $\C_{NS}$ and $\C_{NV}$ in (\ref{phiNS}, \ref{phiNV}) are
fixed  by the condition (\ref{normda0}) ($\C_{NS} = 25.96;\,\C_{NV} =
22.28$). The more complex form of the distribution amplitude for
the quark-vector diquark configuration implies a smaller mean value of
$x_1$ than obtained with the distribution amplitude for the
quark-scalar diquark configuration. In other words, on the average
the $V$-diquark carries a larger fraction of the proton's
momentum than the $S$-diquark. This feature is in accord with the
current expectation that the $V$-diquark mass is larger than that of
the $S$-diquark. \\
The constants $f_{NS}$ and $f_{NV}$, the values of the two
configuration space wave functions at the origin, have also been fixed
in the previous applications of the quark-diquark models. For these
constants the values $f_{NS} = 73.85$ MeV and $f_{NV} = 127.7$ MeV
have been obtained. For the calculation of the $N$--$N^*$ transition
form factors we use the distribution amplitudes (\ref{phiNS}) and
(\ref{phiNV}) with the quoted values of the various parameters. \\
For the $N^*$ the situation is much more complex. On the one hand the
same two types of diquarks, $S$ and $V$ contribute to a $N^*$
state. This requires one unit of orbital angular momentum between
quark and diquark. For these two configurations the appropriate
flavour functions are identical to those of the nucleon
(Eq.~(\ref{chinuc})). The zeroth order terms of the $p$-wave covariant
spin wave functions read
\footnote{We note that $\Gamma_{N^*V}^{\mu\alpha}$ corresponds to the
following case in the ls coupling scheme: The spins of quark and
vector diquark are coupled in a state of total spin 1/2; in the second
step the total spin is coupled with the orbital angular momentum. The
other possibility, namely total spin 3/2, leads to the $S_{11}(1700)$
nucleon resonance. The corresponding spin wave function in this case
reads $ \frac{1}{\sqrt{6}}\left[ g^{\mu \alpha} +
                     \gamma\Tt{f}^{\mu}\gamma^{\alpha} \right] \, .$
Note that the state corresponding to $\Gamma_{N^*V}^{\mu\alpha}$ is a
linear superposition of the two $J^P=1/2^-$ states relevant for
$p$-wave heavy baryon states \cite{koe93}.}
\beq
  \Gamma_{N^*S}^{\alpha}\:=\:\gamma^{\alpha}\,\gamma_5 \, ; \hspace{1cm}
  \Gamma_{N^*V}^{\mu\alpha}\:=\:
      -\frac{1}{\sqrt{3}}\,\gamma\Tt{f}^{\mu}\,\gamma^{\alpha} \, ;
   \label{gamS11}
\eeq
In terms of the non-relativistic quark model these configurations
correspond to mixed-symmetric states. As in the case of $\ell\!=\!0$
states discussed in Sect.~2 we also need correction terms $\sim
K^{\beta}$ to the spin wave functions (\ref{gamS11}). As described in
Appendix A we obtain these corrections from expanding the quark
spinors and the diquark polarisation vectors relative to the direction
of the $N^*$. Quarks and diquarks are considered as free, on-shell
particles. We find
\beqa
  \Delta\Gamma_{N^*S}^{\alpha\beta} & = &
     \frac{1}{2 y_1 M_f}\,\gamma_5\,g^{\alpha\beta} \nonumber \\
  \Delta\Gamma_{N^*V}^{\mu\alpha\beta} & = & -\frac{1}{\sqrt{3}}
       \frac{1}{2 y_1 M_f}\,[2g^{\mu\alpha} - (\gamma\Tt{f}^{\mu}
       - 2 \frac{y_1 P_f^{\mu}}{y_2 M_f}) \gamma^{\alpha}]\,
       \gamma^{\beta}\, .
   \label{DelgamS11}
\eeqa
The Lorentz indices $\alpha$ and $\beta$ are to be contracted with
$K^{\alpha}$ and $K^{\beta}$ (see (\ref{exp1})). We stress that the
above correction terms are model dependent contrary to the zeroth
order spin wave functions. \\
On the other hand the orbital angular momentum that is required to
generate the parity of the $N^*$ may also be inside the diquark.
Hence, in this case, there is no orbital angular momentum between
quark and diquark, i.e. we are back to the $\ell\!=\!0$ case. In order
to account for that possibility we have to introduce three new
diquarks which are parity partners of the $S$- and $V$-diquarks, i.e.
a spin 0, isospin 0 diquark ($P$), a spin 1, isospin 0 diquark ($A_0$)
and a spin 1, isospin 1 diquark ($A_1$). In terms of the
non-relativistic quark model the $q\,P$-, $q\,A_0\,$- and
$q\,A_1\,$-configurations correspond to mixed  antisymmetric states.
As will be discussed in Sect.~4 two of these configurations, namely
$qP$ and $qA_1$, do not contribute to $N$--$N^*$ transitions or can be
neglected. Therefore, we write a $N^*(1535)$ state as
\beqa
 \lefteqn{ |\:N^*;\,\vec P_f,\,\lambda_f\,\rangle \:=\: \int dy_1\:
       [ \,f_{N^*S}\,\phi_{N^*S}(y_1)\,\chi_{N^*S}\,I_{\alpha\beta}/2\,
        (\Gamma^{\alpha}_{N^*S}\:+\:\Delta\Gamma_{N^*S}^{\alpha\beta})
        } \nonumber \\
 & &   \:-\: f_{N^*V}\,\phi_{N^*V}(y_1)\,\chi_{N^*V}\,I_{\alpha\beta}/2\,
        (\Gamma_{N^*V}^{\mu\alpha}\:+\:
         \Delta\Gamma_{N^*V}^{\mu\alpha\beta})\nonumber \\
 & &   \:+\: f_{N^*A_0}\,\phi_{N^*A_0}(y_1)\,\chi_{N^*A_0}\,
         \Gamma_{N^*A_0}^{\mu}\, ] \: u_{N^*}(P_f,\lambda_f) \: ,
         \hspace{2cm}
   \label{S11state}
\eeqa
where we have made use of the definitions
(\ref{intform})-(\ref{normda1}). Obviously, $\chi_{N^*A_0}=\chi_{NS}$
and $\Gamma_{N^*A_0} = \Gamma_{NV}$. \\
In the ansatz (\ref{S11state}) there are three new, a priori unknown
distribution amplitudes and three constants $f_{N^*j}$. To further
simplify the model and to reduce the number of free parameters, we
assume that the $N^*$ distribution amplitudes are of the same form as
the corresponding nucleon distribution amplitudes, (\ref{phiNS}) and
(\ref{phiNV}), for configurations containing a diquark of given spin. The
three constants $f_{N^*j}$ are considered as free parameters to be
determined by fits to experimental data.\\
For the $N^*$ there are two different oscillator parameters,
$\beta_{N^*_0}$ and $\beta_{N^*_1}$, where the indices 0 and 1
corresponding to the two possible values of the orbital angular momentum
$\ell$ between quark and diquark. For the $qA_0$-configuration ($\ell
= 0$) we assume a $k\T$-dependence of the full wave function as in
Eq.~(\ref{ktdep}), and for $N^*$-configurations with a $S$- or
$V$-diquark we assume a $k\T$-dependence as
\beq
   \sim\:k\T\:\exp\left[\,-\beta_{N^*_1}^2\,\frac{k\T^2}{x_1 x_2}\,\right] \, ,
   \label{ktdep1}
\eeq
The ansatz (\ref{ktdep1}) takes care of the fact that we deal with a
$\ell\!=\!1$ system. Requiring again a value of 600 MeV for the rms
transverse momentum, we find a value of 0.71 GeV$^{-1}$ for the
$\ell\!=\!1$ $N^*$ oscillator parameter $\beta_{N^*_1}$. The
corresponding normalization constants of the distribution amplitudes
are $\C_{N^*S}=33.87$ and $\C_{N^*V}=28.65$. For the $\ell\!=\!0$
$N^*$ oscillator parameter $\beta_{N^*_0}$ we have 0.48 GeV$^{-1}$ and
the normalization constant $\C_{N^*A_0}$ is 31.97. For the
constituent masses appearing in the distribution amplitudes (see
Eqs.~(\ref{phiNS}) and (\ref{phiNV})) we take 330 MeV for the quark
and 580 MeV for the $S$- and $V$-diquark as in the nucleon case and
1260 MeV for the $A_0$-diquark. The latter mass should be similar to
that of the $a_1$ meson.

\section{The calculation of the $N$--$N^*$ transition form factors}
\setcounter{equation}{0}

We denote the momentum, the helicity and the mass of the nucleon by
$P_i $, $\lambda_i$ and $M_i$ and the corresponding quantities for the
$S_{11}$-resonance by $P_f$, $\lambda_f$ and $M_f$ (see Fig.~1).
The photon momentum is
\beq
  q \:=\: P_f \,-\, P_i \, .
  \label{q}
\eeq
The $N$--$N^*$ transition matrix element is decomposed as \cite{dev76}
\beq
   \langle\,N^*;\,\vec P_f,\,\lambda_f \:|\:J^{\mu}\:|\:N;\,\vec P_i,
       \,\lambda_i\,\rangle \:=\: e_0\,\bar
      u_{N^*}\,\left[\,h_+(Q^2)/Q_-\,\CH_+^{\mu}\,+\,
      h_0(Q^2)/Q_-\,\CH_0^{\mu}\,\right]\,u_N\, ,
   \label{helcomp}
\eeq
where the helicity covariants are defined by
\beq
   \CH_+^{\mu} \:=\: i\,\Pslash_f\,\epsilon^{\mu\nu\rho\sigma}\,q_{\nu}
                   \,P_{i\rho}\,\gamma_{\sigma}\,, \qquad\qquad
   \CH_0^{\mu} \:=\:
	(P_i\!\cdot\!q\,q^{\mu}\,-\,q^2\,P_i^{\mu})\,\gamma_5 \: .
   \label{helkov}
\eeq
We have factored out the pseudothreshold factor
\beq
   Q_- \:=\: (M_f\,-\,M_i)^2 \:+\: Q^2 \, ,
   \label{qmin}
\eeq
in order to avoid kinematical singularities in the helicity form
factors at the pseudothreshold ($Q_- = 0$).\\
If $\epsilon^{\mu}$ is the polarisation vector of the virtual photon,
then it can easily be seen from (\ref{helcomp}), using for instances
the brick wall frame, that
\beq
   \epsilon_{\mu}(\pm1)\,\CH_0^{\mu} \:=\: \epsilon_{\mu}(0)\,\CH_+^{\mu}
     \:=\: 0 \, .
   \label{helprop}
\eeq
Hence, the helicity form factor $h_+$ refers to transverse and and the
form factor $h_0$ to longitudinal $N$--$N^*$ transitions. While the
covariants (\ref{helkov}) are manifestly gauge invariant, they are not
free of kinematical constraints. At the pseudothreshold one has the
following relation between the helicity form factors
\beq
   h_+(\,Q^2\!=\!-(M_f\!-\!M_i)^2\,)\:=\: \frac{M_i-M_f}{M_f}\:
                            h_0\,(\,Q^2\!=\!-(M_f\!-\!M_i)^2\,)\, .
   \label{helpseudo}
\eeq
In our variant of the hard scattering picture the $N$--$N^*$ transition
matrix elements are given by
\beqa
  \langle\,N^*;\,\vec P_f,\,\lambda_f \:|\:J^{\mu}\:|
      \:N;\,\vec P_i, \,\lambda_i \, \rangle \:=\:
      \bar u_{N^*}(P_f,\,\lambda_f)\,\int\,dx_1 dy_1 \bigg\{
      \hspace{5.5cm}\nonumber \\
   \bigg[\,f_{N^*S}\,\phi^*_{N^*S}(y_1)\,\chi^*_{N^*S}\,
      \frac{I_{\alpha\beta}}{2}\,
      \left(\bar\Gamma^{\alpha}_{N^*S}\,\dd{T_{HSS}^{\mu}}{K'_{\beta}}
        \bigg|_{K'=0}\:+\:\Delta\bar\Gamma_{N^*S}^{\alpha\beta}\,
         T_{HSS}^{\mu}\right) \hspace{4.5cm} \nonumber \\
   +\:f_{N^*A_0}\,\phi^*_{N^*A_0}(y_1)\,\chi^*_{N^*A_0}\,
         \bar\Gamma_{N^*A_0}^{\rho}\,T_{HSA_0\rho}^{\mu}\bigg]\,
         \chi_{NS}\,\Gamma_{NS}\,f_{NS}\,\phi_{NS}(x_1)
         \nonumber \\
   + \,f_{N^*V} \phi^*_{N^*V}(y_1) \chi^*_{N^*V} \frac{I_{\alpha\beta}}{2}
      \left(\bar\Gamma^{\alpha\rho}_{N^*V} \dd{T_{HVV\rho\sigma}^{\mu}}
        {K'_{\beta}}\bigg|_{K'=0}\,+\,\Delta\bar\Gamma_{N^*V}
        ^{\alpha\beta\rho} T_{HVV\rho\sigma}^{\mu}\right)
         \chi_{NV} \Gamma^{\sigma}_{NV} f_{NV}\,\phi_{NV}(x_1)
          \nonumber \\
   \bigg\}\,u_N(P_i,\,\lambda_i)\, \hspace{1cm}
  \label{pS11trans}
\eeqa
where we have made use of the Eqs.~(\ref{0to0da}), (\ref{0to1da}),
(\ref{nucstate}) and (\ref{S11state}). The hard scattering amplitudes
$T_H$ are to be calculated from the Feynman diagrams shown in Fig.~2.
In Eq.~(\ref{pS11trans}) no isospin changing diquark transitions
($S$--$V$ and $V$--$A_0$) appear. For the 3-point
contribution, i.e. for those contributions originating from the
Feynman diagrams where the photon couples to the quark, this fact is
an obvious dynamical property of the model: A gluon cannot mediate
an isospin changing transition. For the 4-point contributions, on the
other hand, where the photon couples to the diquarks, isospin changing
transitions may occur in principle but we make the simplifying
assumption that contributions from such transitions can be neglected.
This assumption is justified by the fact that the 4-point
contributions represent only corrections to the final results. Also in
previous applications of the quark-diquark model
\cite{kss92,kps93,ksP92} such contributions have been neglected.
We mentioned in Sect.~3 that also $qP$ and $qA_1$ configurations
contribute to a $N^*$ state which would lead to additional
diquark-diquark transitions. Yet $S$--$A_1$ and $V$--$P$ transitions
change isospin and therefore do not contribute to $N$--$N^*$
transitions or can be neglected. $S$--$P$ transitions are strictly
zero because of parity.
Furthermore, we have excluded $V$--$A_1$ transitions. This assumption
seems reasonable since, in an analogous calculation of the $\pi$--$a_1$
mesonic transition (see Appendix B), the $V$--$A_1$ transition can
also be shown to vanish asymptotically, when the diquarks are
resolved into quarks.  \\
Thus, we are left with three different transitions, $S$--$S$, $V$--$V$
and $S$--$A_0$, mediated either by a gluon or a photon. The
corresponding vertices for $S$--$S$ and $V$--$V$ transitions, needed in
the calculation of the hard scattering amplitudes have already been
written down in \cite{kss92,kps93,ksP92}. They read
\beqa
 SgS\,: &  \;i\,g_S\,t^a (p_1\,+\,p_2)_{\mu} \hspace{5.5cm}
           \label{SgS} \\
 VgV\,: &  \;-i\,g_S\,t^a
           \bigg[ g_{\alpha\beta} (p_1+p_2)_{\mu}
           \,-\,g_{\mu\alpha}[(1+\kappa_V)p_1 - \kappa_V p_2]_{\beta}
            \nonumber \\
        &  \hspace{4.6cm}-\,g_{\mu\beta}[(1+\kappa_V)p_2 - \kappa_V
           p_1]_{\alpha} \bigg] \, ,
 \label{VgV}
\eeqa
where $g_S\,=\,\sqrt{4\pi\,\alpha_S}$ is the usual QCD coupling
constant which appears here because diquarks are colour antitriplets.
The kinematics used in (\ref{SgS}) and (\ref{VgV}) is defined in
Fig.~3. $\kappa_V$ is the anomalous magnetic moment of the vector diquark
for which we use the value 1.39 \cite{kps93}. $t^a\,=\,(\Lambda^a/2)$
are the Gell-Mann colour matrices. The form of the contact terms
($\gamma SgS$, $\gamma VgV$) required by gauge invariance is obvious.
They can be found in Ref.~\cite{kss91}. The corresponding vertices
with photons instead of gluons are obtained by replacing
$g_S\,t^a$ by $-e_0\,e_D$ ($e_D$ is the charge of the diquark $D$). \\
The new vertex which appears here for the process under investigation
is the spin 0 - gluon (photon) - spin 1 diquark vertex $Sg(\gamma)A_0$.
Its structure can be taken from the analogous $\pi$--$a_1$ transition
current (\ref{pia1decomp}). There are two covariants in general and
correspondingly two form factors. Their asymptotic behaviours are given
by (\ref{f1pia1}) and (\ref{f2pia1}). Our numerical estimates of these
form factors for the $\pi$--$a_1$ case provide the approximate relation
$F_1 \approx Q^2 F_2$. Assuming that a similar relationship holds for
the $S$--$A_0$ case we can write down, in the same spirit as we
constructed the other diquark vertices, a simple gauge invariant
vertex for $S$--$A_0$ transitions
\beq
   SgA_0\,: \;i\,g_S\,\frac{t^a}{m_{S}}\,
      \left[ q^2\,g_{\mu\alpha}\,-\,q_{\mu}q_{\alpha}\right]\:.
   \label{SAvertex}
\eeq
For on-shell diquarks this vertex reduces to (\ref{pia1decomp}) with
$F_1=Q^2 F_2$. As it will turn out the $S$--$A_0$ contributions
only provide a small correction to the $N$--$N^*$ transition form
factors the ansatz (\ref{SAvertex}) is sufficient. For the same reason
we can also safely neglect $S$--$A_0$ 4-point contributions. \\
In applications of the quark-diquark model Feynman diagrams are
calculated with the rules for point-like particles. However, in order to
take into account the composite nature of the diquarks
phenomenological vertex functions have to be introduced. The 3-point
functions, i.e. the ordinary diquark form factors, are parameterized
such that asymptotically the diquark model evolves into the pure quark
model of Brodsky and Lepage \cite{bro80}. In view of this the diquark
form factors are parameterized as
\beqa
   F_S(Q^2) & = \: F_S^{(3)}(Q^2) & = \: (1 \,+\, Q^2/Q_S^2)^{-1}
   \label{FSQ2} \\
   F_V(Q^2) & = \: F_V^{(3)}(Q^2) & = \: (1 \,+\, Q^2/Q_V^2)^{-2} \, ,
   \label{FVQ2}
\eeqa
and
\beq
   F_{SA}(Q^2) \: =\: F_{SA}^{(3)}(Q^2) \:=\: (1 \,+\, Q^2/Q_{SA}^2)^{-2}
   \, ,\label{FSAQ2}
\eeq
where the zero of the $S$--$A_0$ form factor at threshold is ignored;
we are not interested in the small $Q^2$ region. In accordance with
the correct asymptotic behaviour the 4-point functions are
parameterized as
\beq
   F_S^{(4)}(Q^2) \:=\: a_S\,F_S(Q^2)\: ; \qquad\qquad
   F_V^{(4)}(Q^2) \:=\: a_V\,F_V(Q^2)\, (1 \,+\, Q^2/Q_V^2)^{-1} \, .
   \label{SV4poi}
\eeq
The required asymptotic behaviour of the diquark form factors can be
calculated using the usual hard scattering picture along the same lines
as for meson form factors ($\pi\rightarrow\pi$, $\rho\rightarrow\rho$,
$\pi\rightarrow a_1$; for the latter see Appendix B). The first two
form factors have been successfully applied in previous applications
of the quark-diquark picture \cite{kss92,kps93,kss91}, where also the
form factor parameters entering (\ref{FSQ2})-(\ref{SV4poi}) have been
determined. We take the values from Ref.~\cite{kps93}:
\beq
   Q_S^2 \:=\: 3.22\,\mbox{GeV}^2 \, ; \qquad\qquad
   Q_V^2 \:=\: 1.50\,\mbox{GeV}^2 \, .
   \label{Qpar}
\eeq
For the sake of simplicity we set $Q_{SA}^2=Q_V^2$. The physical
picture underlying this assumption is that $V$- and $A_0$-diquarks
both dissolve into quarks at a momentum scale which is smaller than
that for $S$-diquarks. The constants $a_S$ and $a_V$ in the 4-point
functions are strength parameters, which take into account absorption
due to diquark excitation and break-up. Again we take the values for
$a_S$ and $a_V$ from Ref.~\cite{kps93}: $a_S = 0.15$; $a_V = 0.05$. \\
Having specified the rules for the phenomenological diquark
vertices we are now in the position to calculate the hard scattering
amplitudes and hence the current matrix elements (\ref{pS11trans}).
A comparison with the general covariant decomposition of the current
matrix elements, Eq.~(\ref{helcomp}), yields the helicity amplitudes
of the form factors. Before presenting the results a remark concerning
the $K$-dependence of the hard scattering amplitudes $T_{HSS}$ and
$T_{HVV}$ is in order. Obviously, part of the $K$-dependence
comes from the propagators. According to the kinematics defined in
Fig.~2, the internal quark and gluon momenta that contribute to the
3-point functions given by the two Feynman diagrams read
\beqa
   q_{G} & = & x_2\,P_i \,-\, y_2\,P_f \,+\, K \nonumber \\
   q_f   & = & P_f \,-\, x_2\,P_i \nonumber \\
   q_i   & = & P_i \,-\, y_2\,P_f \,+\, K \, .
   \label{intmomenta}
\eeqa
To a very good approximation the denominators of the corresponding
propagators are
\beqa
   q_{G}^2 & = & - x_2 y_2 Q^2\,+\, K^2 \,+\, \Or(M_i^2,M_f^2)\\
   q_f^2\,-\,m_q^2 & = & - x_2 Q^2 \,+\, \Or(M_i^2,M_f^2) \\
   q_i^2\,-\,m_q^2 & = & - y_2 Q^2 \,+\, K^2
   \,+\, \Or(M_i^2,M_f^2) \, . \label{props}
\eeqa
For the Feynman diagrams contributing to the 4-point functions the
internal momenta and propagator denominators are obtained from the
above ones by the replacement $x_1\leftrightarrow x_2$ ;
$y_1\leftrightarrow y_2$. We see that the denominators depend only on
$K^2$. Hence, to the order we are working at, the denominator
$K$-dependence can be neglected. Thus, the $K$-dependence of the hard
scattering amplitude originates from the coupling of photons and/or
gluons to the diquarks and from the quark (and diquark) propagator
numerators related to the momentum $q_i$. From these terms we have to
compute the derivative of the hard scattering amplitude with respect
to $K$.

\section{Results and comparison with data}
\setcounter{equation}{0}

For a qualitative discussion of our results we first write down the
analytical expressions for the $S$--$S$ contributions to the
$p\,$--$S_{11}(1535)$ helicity form factors. These are obtained by
working out (\ref{pS11trans}) and then comparing the result with the
general structure (\ref{helcomp}). One has
\beqa
  \lefteqn{h_+^{S-S}(Q^2) \:=\:-C_F\,\frac{8\pi}{3 Q^2}\,
        \frac{f_{N^*S}}{M_f^2}\,f_{NS}\,\int_0^1\,dx_1dy_1\:
        \phi_{N^*S}(y_1)\,\frac{1}{y_1}}  \nonumber \\
  & &   \times\: \left[\,2\,\fra{\alpha_S(\tilde{Q}_{22}^2)}{x_2 y_2}
	\,F_S^{(3)}(\tilde{Q}_{22}^2) \:+\:
        \fra{\alpha_S(\tilde{Q}_{11}^2)}{x_1 y_1}\,
        F_S^{(4)}(\tilde{Q}_{11}^2+\tilde{Q}_{22}^2)\,\right]\:
        \phi_{NS}(x_1)
	\label{hplSS} \\
\vspace{0.5cm}
  \lefteqn{h_0^{S-S}\,(Q^2) \:=\: -C_F\,\frac{8\pi}{3 Q^4}\,f_{N^*S}\,f_{NS}
        \,\int_0^1\,dx_1 dy_1\:\phi_{N^*S}(y_1)}  \nonumber \\
  & &   \times\: \Bigg[\,2\,\fra{\alpha_S(\tilde{Q}_{22}^2)}{x_2 y_2}
        \,F_S^{(3)}(\tilde{Q}_{22}^2)\,\bigg(1+\frac{x_1 M_i}{y_1 M_f} \bigg)
        \:+\: \fra{\alpha_S(\tilde{Q}_{11}^2)} {x_1^2 y_1^2}\,
        F_S^{(4)}(\tilde{Q}_{11}^2+\tilde{Q}_{22}^2)\, \nonumber \\
  & & 	\times\:\frac{1}{2}\,\left(\,
        \left(1\,+\,\frac{x_1 M_i}{y_1 M_f}\right)(1+x_1 y_1 + x_2 y_2)
        \,-\,4 \right) \Bigg]\:
        \phi_{NS}(x_1) \, , \hspace{5cm}
        \label{hzeSS}
\eeqa
where $\tilde{Q}_{ij}^2 \:\equiv\:x_i\,y_j\,Q^2$ .
Contrary to the case of elastic proton form factors the $S$ diquarks
contribute to the helicity changing $N$-$N^*$ transitions ($h_0^{S-S}$).
The reason for this is obvious: hadron helicity flip contributions can
be mediated via the orbital angular momentum of the $N^*$ without
changing the quark helicity. \\
As discussed in Sect.~2 the $\ell\!=\!0 \rightarrow \ell\!=\!1$
transition current matrix element is given by the sum of two parts:
The first one contains the $\ell\!=\!1$ wave function to
first order in $k\T$ and the derivative of the hard scattering
amplitude with respect to $k\T$, and the second one contains the
correction $\Delta\Gamma$ and the hard scattering amplitude to zeroth
order in $k\T$. It is to be noted that only terms from the second part
$\sim \Delta\Gamma$ contribute to $h_+$, whereas $h_0$ is
determined by both parts. Furthermore, we note that $h_+^{S-S} \sim
Q^{-4}$ and $h_0^{S-S} \sim Q^{-6}$ asymptotically (when use is made
of Eqs.~(\ref{FSQ2}) and (\ref{SV4poi})), i.e. the transverse form
factor dominates for large $Q^2$ as it should. \\
For the $V$--$V$ contributions we obtain
\beqa
  \lefteqn{h_+^{V-V}(Q^2) \:=\:
        -C_F\,\frac{\pi}{3}\,f_{N^*V}\,f_{NV}\,\int_0^1\,dx_1dy_1\:
	\phi_{N^*V}(y_1)\,
	\fra{\alpha_S(\tilde{Q}_{11}^2)}{x_1 y_1 M_f^2 m_V^2}\,
        F_S^{(4)}(\tilde{Q}_{11}^2+\tilde{Q}_{22}^2) } \; \nonumber \\
  & &  \times \,\Bigg[\frac{\kappa_V}{y_2}
	\bigg((1+x_1)\kappa_V + 2x_1\bigg)
        \bigg(3 y_1 - 1 + (1+y_1) \frac{x_1 M_i}{y_1 M_f} \bigg)
        + \frac{y_1 M_f}{M_i}\kappa_V(\kappa_V-1)  \nonumber \\
  & &  +\,(\kappa_V-1)^2(2+3x_1)-2\kappa_V+\frac{3-\kappa_V}{y_1}\,
        (x_2 + y_1 - 4 x_1 y_1 ) \Bigg] \,\phi_{NV}(x_1)
        \label{hplVV} \\ \nonumber \\
\vspace{0.5cm}
  \lefteqn{h_0^{V-V}\,(Q^2) \:=\: -C_F\,\frac{\pi}{3}\,
	f_{N^*V}\,f_{NV}\,\int_0^1\,dx_1dy_1\:
	\phi_{N^*V}(y_1)\,
	\fra{\alpha_S(\tilde{Q}_{11}^2)}{x_1 y_1 y_2 M_i M_f m_V^2}\,
         F_S^{(4)}(\tilde{Q}_{11}^2+\tilde{Q}_{22}^2)\,} \; \nonumber \\
  & &  \times \,\Bigg[\kappa_V(1+y_1)\left(
        \frac{M_i}{y_1 M_f}\left[2(1+x_1)(\kappa_V-1)+3x_1\right]\,
        -\,1\right) + 2 y_2 (1 + \kappa_V)^2 \Bigg]
	\phi_{NV}(x_1) \, . \hspace{1.0cm}  \label{hzeVV}
\eeqa
Because of charge cancellation there are no 3-point $V$--$V$
contributions to $p\rightarrow S_{11}(1535)$ like in the case of
elastic proton form factors. Contrary to $h_+^{S-S}$, the leading
order of $h_+^{V-V}$ also receives contribution from terms involving
derivatives of the hard scattering amplitude with respect to $k\T$.
The major difference, however, is that asymptotically $h_+^{V-V}\sim
Q^{-6}$, i.e. $h_+^{V-V}$ is subdominant in its large $Q^2$ behaviour. \\
The contributions from $S$--$A_0$ diquark transitions read
\beqa
  \lefteqn{
        h_+^{S-A}(Q^2) \:=\ -C_F\,\frac{16\pi}{3\sqrt{3}}\,
	\frac{f_{N^*A_0}f_{NS}}{m_S}\,\int_0^1\,dx_1 dy_1\:
        \phi_{N^*A_0}(y_1) \,\fra{\alpha_S(\tilde{Q}_{22}^2)}{y_2\,M_f^2}
        \,F_{SA}^{(3)}(\tilde{Q}_{22}^2) \:\phi_{NS}(x_1) }
        \label{hplSA}  \\
\vspace{0.5cm}
  \lefteqn{
        h_0^{S-A}\,(Q^2) \:=\: C_F\,\frac{16\pi}{3\sqrt{3}}\,
	\frac{f_{N^*A_0}f_{NS}}{m_S}\,\int_0^1\,dx_1dy_1\:
        \phi_{N^*A_0}(y_1) \,  }  \nonumber \\
   & &  \times \,
        \fra{\alpha_S(\tilde{Q}_{22}^2)}{y_2 Q^2}\,
        F_{SA}^{(3)}(\tilde{Q}_{22}^2)\,
        \left(5 y_2 - 1 - \frac{x_1 M_i}{M_f} \right)\:
        \phi_{NS}(x_1) \, .  \hspace{6cm}
        \label{hzeSA}
\eeqa
Note that $h_+^{S-A}$ and $h_0^{S-A}$ behave as $Q^{-6}$
asymptotically as in the $V$--$V$ case.  \\
Throughout the running coupling constant is taken to be
$\alpha_S(Q^2) = 12\,\pi\,/\,(25\,\ln(Q^2/\Lambda_{\scr{QCD}}^2))$
with $\Lambda_{\mbox{\scriptsize QCD}}=200$ MeV. In order to avoid
singularities for $Q^2 \rightarrow \Lambda_{\mbox{\scriptsize QCD}}^2$
we cut off $\alpha_S$ at a value of 0.5. Keeping in mind that our model
relies on perturbative methods with the implicit assumption that the
running strong coupling constant $\alpha_S(Q^2)$ is small we do not
believe our model to be applicable below $Q^2\lsim 4$ GeV$^2$. \\
In the expressions (\ref{hplSS})--(\ref{hzeSA}) there are a few free
parameters, namely $f_{N^*S}$ and $f_{N^*V}$, the ``derivatives of the
$N^*$ wave function at the origin of the configuration space'' and
$f_{N^*A_0}$. The DAs and other parameters appearing in these
expressions are taken from the study of the electromagnetic nucleon
form factors \cite{kss91}. We introduce a further simplification in
this first explorative investigation of $\ell\!=\!0 \rightarrow
\ell\!=\!1$ transitions by assuming $f_{N^*S} = f_{N^*V}$. Thus,
altogether we are left with two parameters to be fitted to the
experimental data. \\
In \cite{sto91} a compilation is given of both exclusive
and inclusive data on the transition $p\rightarrow S_{11}(1535)$ (see
caption to Fig.~4 for details). The exclusive data, however,
is limited to values of $Q^2 \le 3$ GeV$^2$, whereas the inclusive
data covers a range in $Q^2$ up to about 20 GeV$^2$. It is
given in terms of the helicitity amplitude $A_{1/2}$ which is
proportional to the helicity conserving form factor $h_+$
\cite{kon90,lyt78,war90}
\beqa
   A_{1/2} &\:=\:& \sqrt{\frac{M_f}{M_f^2-M_i^2}}\:
           \langle\,N^*,\,1/2 \:|\:\epsilon_{\mu}(1)\,J^{\mu}\:|\:
           N,\,1/2\,\rangle\nonumber \\
           &\:=\:& - e_0\,M_f\,\sqrt{\frac{Q_+}
                 {8 M_i\,(M_f^2-M_i^2)}}\:h_+(Q^2) \: ,
   \label{helamp}
\eeqa
where $Q_+\,=\,(M_f + M_i)^2\,-\,q^2$. For $Q^2 \ge 3$ GeV$^2$ the
resonance cross section is dominated by the $S_{11}$ around 1535 MeV
invariant mass. Experimentally $A_{1/2}$ was determined from a
resonance fit to the cross section where the assumption was made that
contributions from helicity changing transitions to the cross section
can be neglected (see Ref.~\cite{sto91} for details on the amplitude
extraction procedure). \\
Frequently one uses also the magnetic transition form factor $G_M$
which, in analogy to the Sachs form factors of the nucleon, is defined
by
\beq
  G_M\:=\:-\frac{M_f}{2} \sqrt{\frac{Q_+}{Q_-}}\,h_+ \, .
  \label{sachs}
\eeq
In Fig.~4 we show the data on $G_M$ \cite{sto91,bra76,hai79} and our
model results with $f_{N^*S} = f_{N^*V} = 0.08\:\mbox{GeV}^2$ and
$f_{N^*A_0}=0.05\:\mbox{GeV}$. For $Q^2$ larger than about 6 GeV$^2$
$G_M$ behaves as $Q^{-4}$ as is predicted by the hard scattering
picture of Brodsky and Lepage \cite{bro80}. Our values for $G_M$ (or
$A_{1/2}$) are much larger than those obtained by Carlson and Poor
\cite{car88} who used the Brodsky-Lepage model and the QCD sum rule
constrained Chernyak-Zhitnitsky DA \cite{che89}. As already mentioned
in the introduction Carlson and Poor's ansatz does not treat the
orbital angular momentum of the $S_{11}$ properly. For comparison we
also display the results obtained by Konen and Weber \cite{kon90} in
Fig.~4. Because the Konen-Weber model is a relativistically
generalized constituent quark model which does not account for hard
processes, its validity is limited to the low $Q^2$ regime ($\lsim$ 3
GeV$^2$). \\
In Fig.~5 we plot the helicity form factors $h_+$ and $h_0$
(multiplied by $Q^4$) and their decomposition into the contributions
from the various diquark transitions
(Eqs.(\ref{hplSS})-(\ref{hzeSA})). We observe that the dominant
contribution to $h_+$ is provided by the $S$--$S$ contribution. The
$V$--$V$ and $S$--$A_0$ transitions only supply very small corrections
to $h_+$ which could even be ignored without degrading the quality of
the fit. This result justifies our crude treatment of the $S$--$A_0$
transitions which represents a new piece in the diquark model.
Obviously, since the $S$--$A_0$ contribution is so small, the
parameter $f_{N^*A_0}$ is not well constrained by the data. For the
other helicity form factor $h_0$ the situation is quite different. The
contributions from the $V$--$V$ and the $S$--$A_0$  transitions are
not negligible as compared to those from the $S$--$S$ transitions.
Data on that form factor would pin down the magnitude of the
$S$--$A_0$ contributions. There are a few data points on $h_0$ in the
2--3 GeV$^2$ region \cite{loq2data}. Fig.~5 shows that the trend
of our results is in fair agreement with the data on $h_0$.

\section{Summary and conclusions}

In this paper we have proposed a method to extend the hard scattering
picture of Brodsky and Lepage \cite{bro80} to hadrons with nonzero
orbital angular momentum where one can no longer neglect the internal
transverse momentum. We have applied this formalism to the $N$--$N^*$
transition in the quark-diquark picture and to the mesonic transition
$\pi \rightarrow a_1$. The generalization of our approach to baryon
reactions in the pure quark picture is straightforward and will be
performed in a later paper. We emphasize that our formalism as well as
the original Brodsky-Lepage approach is only valid at large momentum
transfer. In the few GeV region, however, radiative corrections
\cite{lis92} have to be taken into account as well. In the present
explorative study we have ignored this complication. \\
Concerning possible quark-diquark configurations inside the $N^*$ the
orbital angular momentum can either be between quark and diquark or
reside inside the diquark. In the former case one encounters the
well-known $S$- and $V$-diquarks, whereas the latter case requires the
introduction of new diquarks with new couplings. However, only three
configurations contribute to the $N$--$N^*$ transition, namely the
diquark transitions $S\rightarrow S$, $V\rightarrow V$ and the new
transition $S\rightarrow A_0$. In the numerical analysis we found that
the contributions from $S$--$S$ diquark transitions provide the dominant
part of the non-flip helicity form factor $h_+$, whereas the behaviour
of the helicity flip form factor $h_0$ is mainly governed by $V$--$V$
and $S$--$A_0$ transitions. \\
It is important to note that the phenomenology of this treatment is
quite different from that of Refs.~\cite{car88,kss92}, where the $N^*$
was treated uncorrectly as a nucleon with opposite parity. Although
our model of the $N^*$ involves some unknown parameters we believe
that our approach constitutes a step forward towards an understanding
of hadrons with nonzero orbital angular momentum.

\section*{Acknowledgements}

We would like to thank R.~Jakob, M.~Sch\"urmann and W.~Schweiger for
many useful discussions. One of us (J.~B.) acknowledges a
graduate scholarship of the Deutsche Forschungsgemeinschaft.


\begin{appendix}

\renewcommand{\thesection}{Appendix \Alph{section}:}
\renewcommand{\theequation}{\Alph{section}.\arabic{equation}}

\setcounter{equation}{0}
\section{Derivation of the spin wave functions for the $N$(1/2$^+$)
and the $N^*$(1/2$^-$) in the $ls$ coupling scheme}

One way of constructing covariant spin wave functions is to
make use of the observation \cite{dzi88} that, in the zero binding energy
approximation, an equal-$t$ hadron state in the constituent center of mass
frame ($\sum \vec{\hat{p}}_j = 0$) equals the light cone state at rest.
Consequently, one can use the standard $ls$ coupling scheme to couple
quarks and diquarks in the baryon case, or quarks and antiquarks in
the meson case to form a state of given spin and parity. On boosting
the result to a frame with arbitrary hadron momentum $P^{\mu}$ one can
easily read off the covariant spin wave function. As was mentioned in
Sect.~2 one needs $K^{\alpha}$-corrections to the spin wave functions for
$\ell\!\not=\!0$ hadrons. As we will see below the approach presented
in this appendix also provides a scheme for calculating such
corrections which admittedly is model-dependent. \\
To be specific we will describe the construction of covariant spin
wave functions in the $ls$ coupling scheme for spin 1/2 baryons made
of quarks and diquarks of type $j$ with spin $s_j$ and with a relative
orbital angular momentum $\ell$ between quark and diquark. The
generalization to other cases is straightforward. The center of mass
momenta are denoted by
\beq
  \hat P^{\mu}   = (M, \vec 0)   \, ; \qquad\qquad
  \hat p_1^{\mu} = (m_1, \vec k) \, ; \qquad\qquad
  \hat p_2^{\mu} = (m_2, -\vec k)\, ,
  \label{cmsmomenta}
\eeq
and, according to the discussion in Sect.~2, we have the approximate
relations $m_i = x_i M + \Or(k^2/M),\:i\!=\!1,2$. The baryon is
represented by the equal-$t$ spinor $u_B(\hat P,\mu)_t$ and its
constituents, the quark and the diquark $j$, by the spinors
$u(\hat p_1, \mu_1)_t$ and by $\hat\epsilon_j(\hat p_2, \mu_2)$ which
is 1 for a scalar diquark and a Lorentz vector for the case $s_j=1$.
Note that $\mu$, $\mu_1$ and $\mu_2$ denote spin components. The $ls$
coupling scheme leads to the following ansatz for the spin wave
function
\beqa
      \hat\Gamma_{j\ell}\,u_B(\hat P, \mu)_t &=& \sum_{\mu_1
      \mu_2 \mu_{\ell}}\:|\,\vec k\,|\,^{\ell}\,\sqrt{4\pi}\,
      Y_{\ell \mu_{\ell}}(\vec k /\,|\vec k|\,)\, \nonumber \\
  & \times & \bma{cc|c}
       1/2 & s_j & s \\ \mu_1 & \mu_2 & \mu_s
      \ema\,
      \bma{cc|c}
        s & \ell & 1/2 \\ \mu_s & \mu_{\ell} & \mu
      \ema
      \,u(\hat p_1, \mu_1)_t \, \hat\epsilon_j(\hat p_2, \mu_2)\: .
    \label{lsans}
\eeqa
Possible Lorentz indices of the baryon spin wave function and of
$\hat\epsilon_j$ are omitted for the present discussion. The quark
spinor is related to the baryon spinor by
\beq
    u(\hat p_1,\mu_1)_t\:=\: \frac{1}{2 x_1 M} \:
    (x_1\Phslash\,+\,\Khslash\,+\,x_1 M) \:
    u_B(\hat P, \mu_1)_t \, ,
    \label{spinboo}
\eeq
where the 4-vector $\hat K^{\mu} = (0, \vec k)$ has been introduced.
Up to corrections of order $k^2/M^2$ the polarization vector of a
$s_j\!=\!1$ diquark can be expressed as
\beq
    \hat\epsilon_j\,\!^{\nu}(\hat p_2,\mu_2)\:=\:
      \left(g^{\nu}\,\!_{\rho}\:+\:\frac{1}{x_2 M^2}\,\hat P^{\nu}
      \hat K_{\rho}\right)\,\hat\epsilon_j\,\!^{\rho}(\hat P,\mu_2) \, ,
    \label{epsboo}
\eeq
where $\hat \epsilon_j\,\!^{\rho}(\hat P,\mu_2)$ describes the polarization
state of a vector diquark at rest. Thus, inserting (\ref{spinboo}) and
(\ref{epsboo}) in (\ref{lsans}), one can easily find the rest frame
spin wave functions including, for $\ell\!=\!1$, the corrections
proportional to $K^{\alpha}$. Boosting to a frame with baryon momentum
$P^{\mu}\:=\:(E,\,\vec P)$ one arrives at the desired spin wave
functions written down in Sect.~3 (and for the $\pi$ and $a_1$ meson
in Appendix B). They include the corrections $\Delta \Gamma$ obtained
from the expansion of the quark spinor and the polarisation vector
around the direction of the hadron momentum. Since this has been
carried out for free quark spinors respective polarisation vectors the
result for $\Delta\Gamma$ is model dependent in contrast to the
leading terms of the spin wave functions which only depend on the
hadronic quantum numbers and the type of the constituents.

\section{The $\pi$--$a_1$ form factors}
\setcounter{equation}{0}

In this Appendix we present a calculation of the $\pi$--$a_1$ form
factors at large momentum transfer. The reason for doing this is
twofold. On the one hand we want to demonstrate that our approach is
also applicable to mesonic transitions and, on the other hand, we can
read off the asymptotic behaviour of the $S$--$A_0$ diquark form
factors from the results. We need to know the large $Q^2$ power
dependence for the parameterization of the diquark form factors in
order to guarantee that asymptotically the pure quark model emerges
from the quark-diquark picture. \\
We use analogous notations and the same frame as for the calculation
of the $N$--$N^*$ transition form factors (see Sect.~2). We
covariantly decompose the current matrix element as follows
\beqa
  \langle\,a_1;\,\vec P_f,\,\lambda_f \:|\:J^{\mu}\:|
      \:\pi;\,\vec P_i \, \rangle \:= \hspace{11cm} \nonumber \\
   -i\,e_0\,\left[\, F_1(Q^2)\,( q\!\cdot\!P_f\,g^{\mu\nu}
                     \,+\, P_f^{\mu} P_i^{\nu} )
         \:+\: F_2(Q^2)\,( q\!\cdot\!P_f\,P_i^{\mu}
                     \,-\, q\!\cdot\!P_i\,P_f^{\mu} )\,P_i^{\nu}
     \right] \, \epsilon_{\nu}\, .\hspace{1cm}
  \label{pia1decomp}
\eeqa
The covariants are manifestly gauge invariant. Using again the opposite
momentum brick wall frame one concludes that $F_1$ is the form factor
for transverse transitions and, for $Q^2 \rightarrow \infty$, $F_2$ is
the form factor for longitudinal transitions. \\
According to our statements made in Sect.~2 we write the mesonic
states as \cite{hus91}
\beqa
  \label{pistate}
  |\:\pi;\,\vec P_i\,\rangle &\:=\:& \int\,dx_1\,f_{\pi}\,
                           \phi_{\pi}(x_1)\,\Gamma_{\pi} \\
  \label{a1state}
  |\:a_1;\,\vec P_f,\,\lambda_f\,\rangle &\:=\:&
        \int\,dy_1\,f_{a_1}\,\phi_{a_1}(y_1)\,
        (\Gamma_{a_1}^{\alpha}\,+\,\Delta\Gamma_{a_1}^{\alpha\beta})\, ,
\eeqa
where
\beqa
  \label{gampi}
  \Gamma_{\pi} &\:=\:& \frac{1}{\sqrt{2}}\,(\Pslash_i\,+\,M_i)\,\gamma_5 \\
  \label{gama1}
  \Gamma_{a_1}^{\alpha} &\:=\:& \frac{1}{2}\,(\Pslash_f\,+\,M_f)\,
       [\sla{\epsilon}(\lambda_f),\gamma^{\alpha}]\,\gamma_5 \, ,\\
\eeqa
and
\beq
  \Delta\Gamma_{a_1}^{\alpha\beta}\:=\:g(y_1)\,\gamma^{\beta}\,
       [\sla{\epsilon}(\lambda_f),\gamma^{\alpha}]\,\gamma_5 \, .
  \label{delgam}
\eeq
The spin wave function $\Gamma_{a_1}$ (Eq.~(\ref{gama1})) has been
given before in \cite{bhk93,koe93,ji92}. The correction term
$\Delta\Gamma_{a_1}^{\alpha\beta}$ takes into account
that quark and diquark are not collinear with the $a_1$ meson.
The model dependence of this term is absorbed into the function
$g=g(y_1)$. Using the same approach as for the baryon spin wave
functions (see appendix A), we find
\beq
  g(y_1)\:=\: \frac{1}{4 y_1 y_2}\, .
  \label{gy1}
\eeq
We have used the expansion of the free quark spinors around the
direction of the $a_1$ momentum. To lowest order in $\alpha_S$
the hard scattering amplitude is given by (see Eq.~(\ref{0to1da}))
\beqa
   \lefteqn{\langle\,a_1;\,\vec P_f,\,\lambda_f \:|\:J^{\mu}\:|
         \:\pi;\,\vec P_i \, \rangle \:=\: -i e_0 \,\int\,dx_1\,dy_1\:f_{a_1}\,
         \phi^*_{a_1}(y_1)\:\fra{4 \pi\,\alpha_S(q_G^2)\,C_F}{q_G^2}\:
         I_{\alpha\beta}/2} \hspace{0.5cm} \nonumber \\
   & \times & \mbox{Tr}\left\{\left[\bar\Gamma_{a_1}^{\alpha}\,\frac{
         \gamma^{\mu}\gamma^{\beta}\gamma^{\nu}}{q_i^2-m_q^2}\,\:+\:
         \Delta\bar\Gamma_{a_1}^{\alpha\beta}\,\left(
	 \fra{\gamma^{\mu} \sla{q}_i \gamma^{\nu}}{q_i^2-m_q^2}\,\:+\:
	 \fra{\gamma^{\nu} \sla{q}_f \gamma^{\mu}}{q_f^2-m_q^2}\,
	 \right)\right]\,
         \Gamma_{\pi}\,\gamma_{\nu} \right\}\:f_{\pi}\:\phi_{\pi}(x_1)
         \, . \hspace{1cm}
   \label{pia1calc}
\eeqa
In writing Eq.~(\ref{pia1calc}) we have made use of the fact that the
meson wave functions are symmetric under the replacement $x_1
\leftrightarrow x_2$ ($y_1 \leftrightarrow y_2$) due to time reversal
invariance. The propagators are defined in Eq.~(\ref{props}). The
trace can easily be worked out. At large $Q^2$ comparison with
(\ref{pia1decomp}) results in the following expressions for the two
form factors
\beqa
   \label{f1pia1}
   F_1(Q^2) & = & \frac{4\pi\,C_F\,f_{\pi}\,f_{a_1}}{Q^4}\,
                  \int\,dx_1\,dy_1\:\phi^*_{a_1}(y_1)\,\phi_{\pi}(x_1)\,
                  \alpha_S(q_G^2)\:
                  \frac{4}{x_2^2 y_2^2}\,\left(x_2\,+\,2g(1+x_2)y_2
                  \right) \hspace{1cm} \\
   \label{f2pia1}
   F_2(Q^2) & = & \fra{4\pi\,C_F\,f_{\pi}\,f_{a_1}}{Q^6}\,
                  \int\,dx_1\,dy_1\:\phi^*_{a_1}(y_1)\,\phi_{\pi}(x_1)\,
                  \alpha_S(q_G^2)\:
                  \frac{64g}{x_2 y_2} \, ,
\eeqa
where we have neglected the meson masses for convenience. The
additional $Q^{-2}$ factor in (\ref{f2pia1}) is a consequence of the
definitions of the covariants in (\ref{pia1decomp}). $F_2$ is not
suppressed asymptotically in observables relative to $F_1$. For the
model function (\ref{gy1}) the end-point regions,
$x_i\,(y_i)\rightarrow 0,\,i=1,2$, are rather dangerous.  With
distribution amplitudes of the type (\ref{phiNS}) and (\ref{phiNV})
which are damped exponentially in the end-point regions there is,
however, no singularity (except of the familiar difficulty with
$\alpha_S$). When Sudakov corrections are taken into account in the
manner proposed by Sterman et al.~\cite{lis92} the end-point regions
are sufficiently strongly damped for any DA in use. Also the problem
with $\alpha_S$ disappears in this case. \\
With regard to the Sudakov corrections, which are not taken into
account in the present explorative study of $\ell\!=\!0 \rightarrow
\ell\!=\!1$ transitions, we evaluate the $\pi$--$a_1$ transition form
factors only for DAs strongly damped in the end-point regions. We
consider results obtained with other DAs as unreliable overestimates
of the form factors. The following pion DAs are used in the evaluation
of the form factors:  The ``non-relativistic'' DA $\sim
\delta(x_1-1/2)$, the asymptotic form $\sim x_1 x_2$ and the CZ form
$\sim x_1 x_2 (x_1-x_2)^2$; the latter two are multiplied with an
exponential $\exp(-\beta^2 m_q^2 / x_1 x_2)$ \cite{lbh83}. For details
and the values of the oscillator parameter $\beta$ and of the quark
mass $m_q$, see Ref.~\cite{jak93}. The constant $f_{\pi}$ is related
to the pion decay constant by 133 MeV$/2\sqrt{6} = 27.15$ MeV. Not
much is known about the DA of the $a_1$. Ref.~\cite{ji92} contains an
attempt of estimating it. We will, however, not use this result but
rather assume that the DA of the $a_1$ equals the pion's DA. The
constant $f_{a_1}$, which is the derivative of the configuration space
wave function at the origin, can be determined from the decay width
$\Gamma(\tau \rightarrow a_1 \nu_{\tau})$ which equals $\Gamma(\tau
\rightarrow 3\pi \nu_{\tau}) = (1.47 \pm 0.13) \cdot 10^{-10}$ MeV to
a very good approximation \cite{alb93}. The usual decay constant is
defined by $\langle 0 \,|\, J^W_{\mu} \,|\,a_1\rangle =
f_{a_1}^{\scr{dec}} p_{\mu}$ and is related to the constant $f_{a_1}$
by $f_{a_1} = f_{a_1}^{\scr{dec}}/4 \sqrt{3}$. The experimental decay
width leads to $f_{a_1} = 0.025 \pm 0.002 \, \mbox{GeV}^2$. \\
Our numerical results for the two $\pi$--$a_1$ form factors are shown
in Fig.~6. It can be seen that the various choices in the DAs lead to
quite different results for the form factors. The comparatively higher
values obtained from the CZ DAs are due to their stronger
concentration in the end-point regions. The ratio $Q^2 F_1/F_2$,
however, can be found to be equal to a constant of order 1. We
emphasize that we have calculated the large $Q^2$ behaviour of a form
factor, namely $F_1$,  which belongs to a helicity flip current matrix
element. Hadronic helicity is not conserved in this case. The physical
reason is obvious: The hadronic helicity is changed via the orbital
angular momentum; as in the $\ell\!=\!0$ case the quark helicities
remain unchanged in the interaction with photons or gluons (quark
masses are neglected). We mention that the $\pi$--$a_1$ form factors
have been investigated in Ref.~\cite{azn93} at small, time-like
momentum transfer.

\end{appendix}


\section*{Figure captions}

\begin{description}

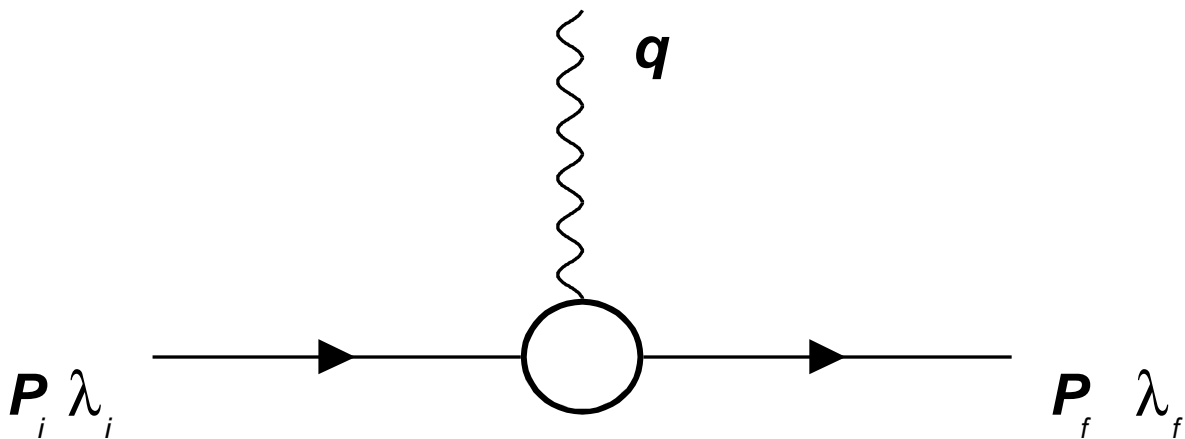
  \item[Fig.~1:]   The $N$--$N^*$ transition vertex.

  \item[Fig.~2a:]  Feynman diagrams contributing to the $N$--$N^*$
	transition form factors to leading order. The upper lines
        represent the quarks, the lower ones the diquarks. The 4-point
	function is drawn schematically as a blob and is explained in
	b) and c).

  \item[Fig.~2b:]  Diagrammatical decomposition of the $pSgS$ and the
	$pVgV$ 4-point function.

  \item[Fig.~3a:]  Diquark-photon/gluon-diquark 3-point vertices.
	$\mu$, $\alpha$ and $\beta$ denote Lorentz indices; $a$ the
	colour of the gluon.

  \item[Fig.~3b:]  Photon-diquark-gluon-diquark 4-point vertices. $q$
	and $k$ denote the momenta of the photon and the gluon,
	respectively.

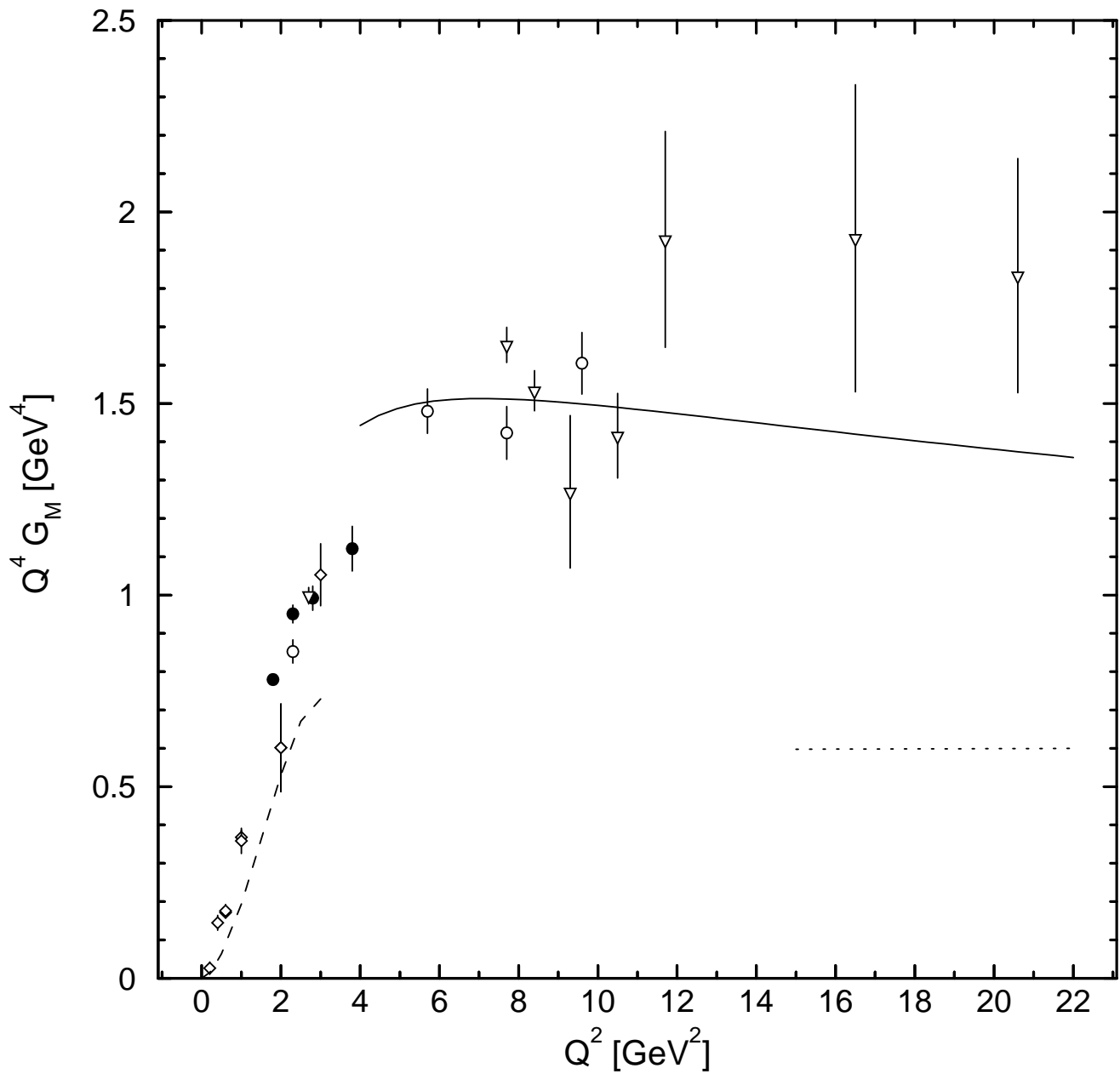
  \item[Fig.~4:]  The magnetic transition form factor $G_M$
	(Eq.~(\ref{sachs})) vs.~$Q^2$. The solid line represents the
	prediction of the diquark model; the dashed line is the result
	of Konen and Weber \cite{kon90} and the dotted line that of
	Carlson and Poor \cite{car88}. Data are taken from
	\cite{sto91} ($\scriptstyle \bigtriangledown$ old, $\circ$ new
	inclusive SLAC data), \cite{bra76} ($\bullet$) and \cite{hai79}
	($\scriptstyle \Diamond$).

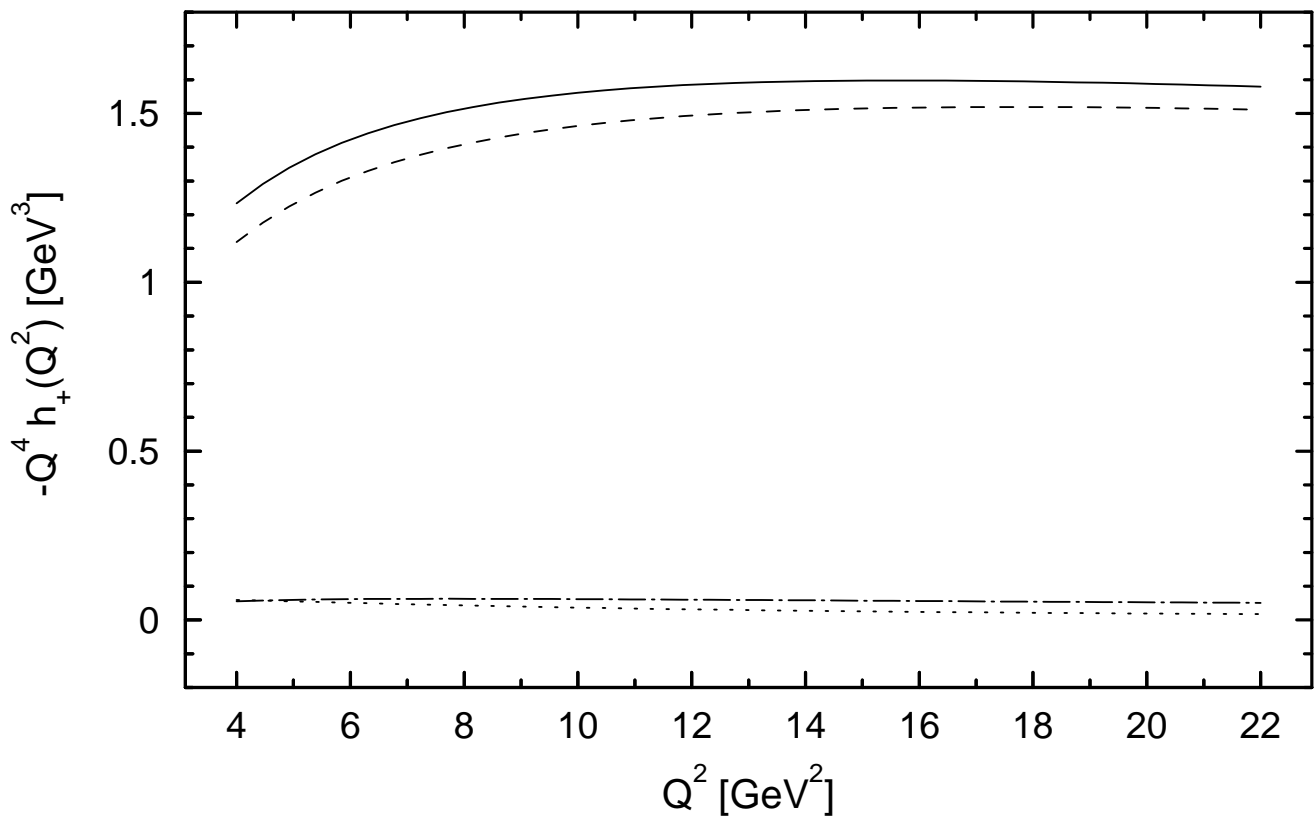
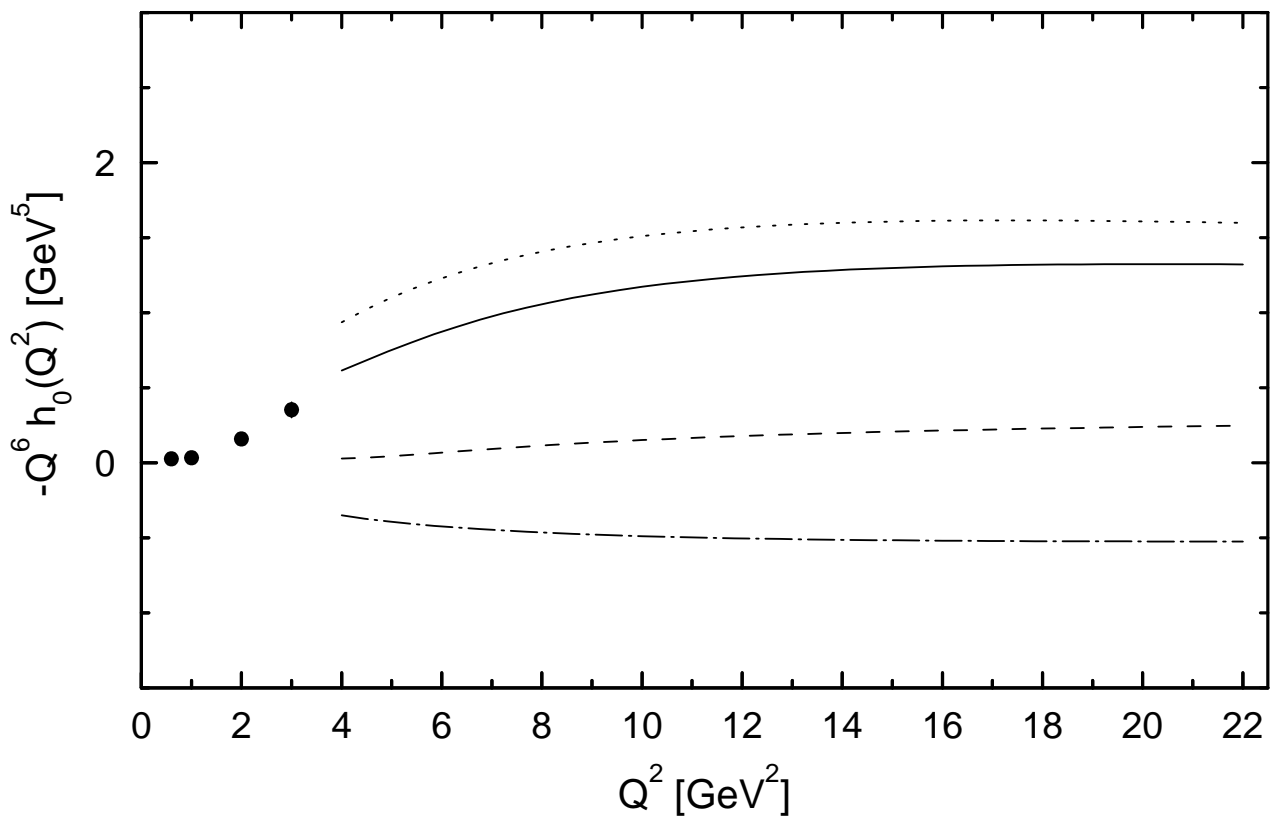
  \item[Fig.~5:]  Helicity form factors $h_+$ (top) and $h_0$ (bottom)
	and their decomposition into contributions from the various
	diquark transitions (Eqs.~(\ref{hplSS})-(\ref{hzeSA})): Dashed
	line are $S$--$S$ transitions, dotted line are $V$--$V$ and
	dashed-dotted line are $S$--$A_0$ transitions. The solid line
	represents the full result.

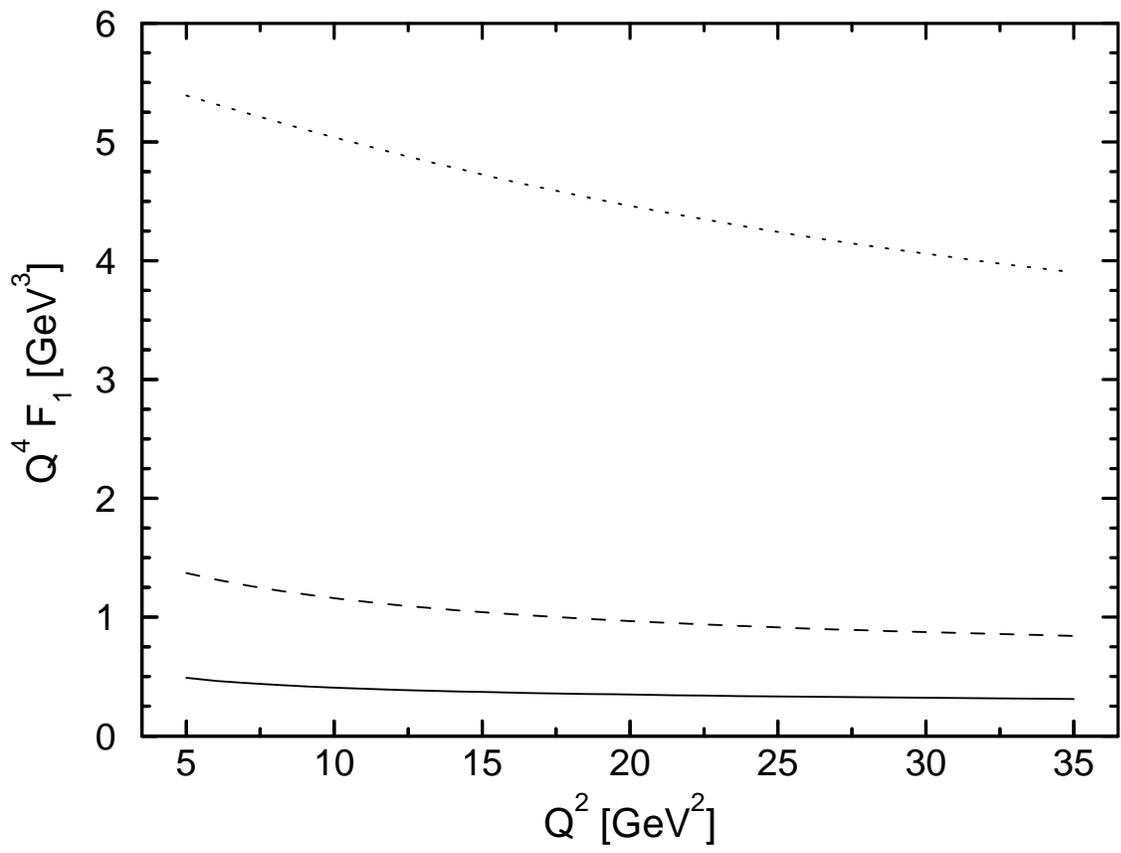
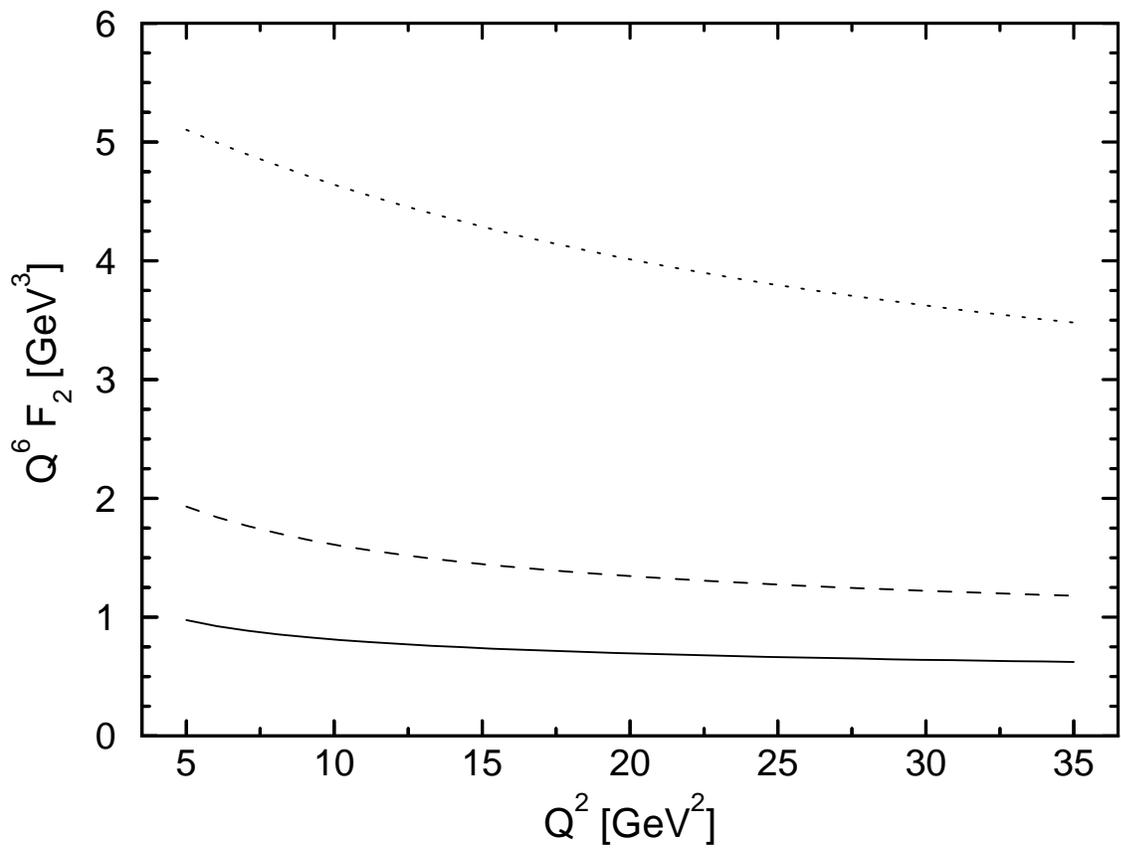
  \item[Fig.~6:]  The $\pi$--$a_1$ transition form factors $F_1$ and $F_2$
	(Eqs.~(\ref{f1pia1}) and (\ref{f2pia1})) vs.~$Q^2$. The lines
	represent our results for the ``non-relativistic'' (solid),
 	the asymptotic (dashed) and the CZ (dotted) DA, assuming that the
 	$\pi$ DA and $a_1$ DA are equal. In all cases the oscillator
	parameter in the DA was taken such as to yield a rms
	transverse momentum of 250 MeV assuming the $k\T$ dependences
	of the complete wave function to be of the form (\ref{ktdep})
	for the $\pi$ and (\ref{ktdep1}) for $a_1$.
\end{description}

\end{document}